%% file: sample.tex
\documentclass[letterpaper, 10 pt, conference]{ieeeconf}  

\IEEEoverridecommandlockouts                              
\usepackage{physics}
\usepackage{amsmath}
\usepackage{bbold}
\usepackage{comment}
\usepackage[english]{babel}
\usepackage{graphicx}
\usepackage[colorlinks=true, allcolors=blue]{hyperref}
\usepackage{booktabs}

\usepackage[dvipsnames]{xcolor}
\usepackage{algorithm}
\usepackage{algorithmic}
\usepackage{booktabs}
\usepackage{bm}
\usepackage{tikz}
\usetikzlibrary{positioning}
\usepackage{amssymb}
\usepackage{subcaption}
\usepackage{mathtools}
\usepackage{enumerate}
\definecolor{darkpowderblue}{rgb}{0.0, 0.2, 0.7}
\definecolor{color1}{HTML}{5ec962} 
\definecolor{color2}{HTML}{21918c} 
\definecolor{color3}{HTML}{3b528b} 
\usepackage{hyperref}
\hypersetup{
    colorlinks=true,
    linkcolor=darkpowderblue,
    linktocpage=true,
    filecolor=magenta,
    citecolor=darkpowderblue,
    urlcolor=cyan
    }
\usepackage{pifont}
\usepackage{placeins}
\usepackage{cuted} 

\title{\LARGE \bf
Variational quantum algorithm for solving Helmholtz problems with high order finite elements
}

\author{Arnaud Rémi, François Damanet and Christophe Geuzaine
\thanks{This work was funded in part by the European Regional Development Fund (ERDF) and the Walloon Region through project 1149 VirtualLab\_ULiege (program 2021-2027).}
\thanks{A. Rémi and C. Geuzaine are with the Dept. of Electrical Engineering and Computer Science, Montefiore Institute, University of Liège, Liège, Belgium
        {\tt\small \{aremi,cgeuzaine\}@uliege.be}}%
\thanks{F. Damanet is with the Institut de Physique Nucléire, Atomique et de Spectroscopie, CESAM, University of Liège, Liège, Belgium
        {\tt\small fdamanet@uliege.be}}%
}

\begin{document}

\maketitle

\begin{abstract}
Discretizing Helmholtz problems via finite elements yields linear systems whose efficient solution remains a major challenge for classical computation. In this paper, we investigate how variational quantum algorithms could address this challenge. 
We first show that, for regular meshes, a block encoding of the operators $A$ and $A^\dagger A$ arising from the high-order finite element discretisation of Helmholtz problems can be designed, resulting in a quantum circuit of depth $\mathcal{O}(p^3\mathrm{poly}\log(Np))$ with $N$ the number of elements and $p$ the order of the finite elements. Then we apply our algorithm to a one-dimensional Helmholtz problem with Dirichlet and Neumann boundary conditions for various wavenumbers.
\end{abstract}

\section{Introduction}

Over the last few decades, quantum computing has gained significant interest due to its potential in solving problems that are classically intractable~\cite{Pfaendler2024}. This growing interest has driven parallel advancements in both hardware architectures and quantum algorithms. Although current quantum hardware remains in the Noisy Intermediate-Scale Quantum (NISQ) era~\cite{Lau2022}, characterized by a limited number of qubits and decoherence, there is a strong interest in developing algorithms that will be effective on the next generation of fault-tolerant quantum computers.

Quantum algorithms have already been established for a wide variety of computational problems. While early applications focused on quantum simulation~\cite{lloyd1996universal, Berry2007, Berry2015, Berry2020, Childs2021}, the field has expanded to include many applications, including algorithms for systems of linear equations~\cite{Ambainis2012, Clader2013, Childs2017, Wossnig2018, Subasi2019, Gilyen2019, Lin2020, Costa2022, Dalzell2024} and, more recently, partial differential equations (PDEs)~\cite{Berry2014, Montanaro2016, Costa2019, Lloyd2020, jin2022quantum, jin2023quantum}. Within this landscape, Variational Quantum Algorithms (VQAs) have garnered particular attention. By leveraging a hybrid classical-quantum optimization framework, VQAs allow for the use of relatively short quantum circuits, making them potential candidates for achieving advantages on near-term hardware.

VQAs have been adapted to solve a range of fundamental engineering problems, including eigenvalue problems~\cite{peruzzo2014variational} and linear systems~\cite{bravo2023variational}. Recent literature has focused on solving linear problems arising from the low-order finite difference or finite element discretization of the Poisson problem~\cite{liu2021variational, sato2021variational}. Furthermore, the application of VQAs to wave propagation has been explored, notably for solving the eigenvalue problem arising from the finite difference discretization of the homogeneous Helmholtz problem~\cite{ewe2022variational}. 

In this paper, we extend these works by proposing a variational quantum algorithm designed to solve non-homogeneous Helmholtz problems using high-order finite elements.

\section{Helmholtz problem formulation}

\subsection{One-dimensional Helmholtz problem}

We consider the following one-dimensional Helmholtz problem on the interval $\Omega=\{x: 0<x<1\}$ with homogeneous Dirichlet and Neumann boundary conditions on the left and on the right, respectively:
\begin{equation}
    \begin{cases}
        \phi''(x) + k^2(x) \phi(x) = f(x), \quad x\in \Omega,\\
        \phi(0) = \phi'(1) = 0,
    \end{cases}
    \label{eq:hermitian_problem}
\end{equation}
where the wavenumber $k(x)=\omega/c(x)$ is defined in terms of the wave velocity $c(x)$ and the angular frequency $\omega$. A discrete form of \eqref{eq:hermitian_problem} can be obtained using the finite element method.

\subsubsection{Finite element method}
The finite element method combines a weak formulation of the original PDE and a Galerkin projection using suitable basis functions.
\paragraph{Weak formulation} To obtain a weak formulation of \eqref{eq:hermitian_problem}, the PDE is multiplied by test functions $v$ in  $H^1_D:=\{ f(x)\in H^1(\Omega): f(0) = 0\}$ and integrated by parts. It consists in finding $\phi(x)\in H^1_D$ such that
\begin{gather*}
    \int_0^1 \phi''v \,\mathrm{d}x + \int_0^1 k^2\phi v \, \mathrm{d}x = \int_0^1 fv \,\mathrm{d}x\\
\begin{split}
    \Leftrightarrow \quad -\int_0^1 \phi' v' \mathrm{d}x + \phi'(1)v(1) - \phi'(0)v(0) + \int_0^1k^2\phi v\, \mathrm{d}x\\= \int_0^1 fv \,\mathrm{d}x
\end{split}
\end{gather*}
holds for all test functions $v\in H^1_D$.
The boundary terms vanish since $v(0)=0$ and $\phi'(1)=0$, and the weak form simplifies into finding $\phi(x)\in H^1_D$ such that
\begin{equation}
    -\int_0^1 \phi' v' \mathrm{d}x + \int_0^1k^2\phi v\, \mathrm{d}x= \int_0^1 fv \,\mathrm{d}x, \quad\forall v\in H^1_D.
    \label{eq:weak_form}
\end{equation}

\paragraph{Galerkin projection} A Galerkin projection is a projection of the solution of the original PDE onto a finite-dimensional subspace spanned by a family of basis functions $\{\varphi_j(x)\}_{j=0}^{N_{\mathrm{dof}}-1}$. The solution $\phi$ of the Helmholtz problem is, hence, approximated by
\begin{equation}
    \phi(x) \approx \sum_{j=0}^{N_{\mathrm{dof}}-1} \phi_j \varphi_j(x),
    \label{eq:galerkin_proj}
\end{equation}
with $\{\phi_j\}_{j=0}^{N_{\mathrm{dof}}-1}$ a set of unknown coefficients.
By substituting \eqref{eq:galerkin_proj} into \eqref{eq:weak_form}, and by using the same basis functions as test functions, the weak formulation can be written in the following discrete form: find the coefficients $\phi_j$ such that
\begin{multline}
    \sum_{j=0}^{N_{\mathrm{dof}}-1} \int_0^1 (-\varphi_j'\varphi_i')\, \mathrm{d}x \;\phi_j + \sum_{j=0}^{N_{\mathrm{dof}}-1} \int_0^1 k^2 \varphi_j \varphi_i \,\mathrm{d}x \;\phi_j \\= \int_0^1 f\varphi_i \,\mathrm{d}x, \quad\forall i.
\end{multline}
In matrix form we obtain:
\begin{equation}
    (K+M)\bm \phi = \bm f, \quad
    \begin{cases}
        K_{ij} = \int_0^1 (-\varphi_i'\varphi_j') \, \mathrm{d}x,\\
        M_{ij} = \int_0^1 k^2 \varphi_i \varphi_j \,\mathrm{d}x,\\
        f_j = \int_0^1 f\varphi_j(x) \,\mathrm{d}x.
    \end{cases}
    \label{eq:linear_system}
\end{equation}
\paragraph{Finite elements}
The spatial domain is discretized into $N$ non-overlapping adjacent finite elements $\{E_\alpha\}$. Each element is associated with $p+1$ nodes $\{x_0^{(\alpha)}, x_1^{(\alpha)}, \dots, x_{p}^{(\alpha)}\}$, which correspond to interpolation points within the element. The chosen family of basis functions for this work are Lagrange polynomials, defined as:
\begin{equation}
\varphi_j^{(\alpha)}(x) = \prod_{\substack{i=0 \\ i \neq j}}^{p} \frac{x - x_i^{(\alpha)}}{x_j^{(\alpha)} - x_i^{(\alpha)}},
\end{equation}
where $\varphi_j^{(\alpha)}(x)$ is the $j$-th basis function associated with node $x_j^{(\alpha)}$, which satisfies $\varphi_j^{(\alpha)}(x_i^{(\alpha)}) = \delta_{ji}$. Within each element $E_\alpha$, the nodes $x_i^{(\alpha)}$ are the Gauss-Lobatto-Legendre (GLL) nodes. The order $p$ of the basis functions as well as the number of elements $N$ are crucial regarding the discretization errors, but also regarding the system condition number. For sufficiently smooth problems, the discretization error of finite elements with GLL nodes is $\mathcal{O}(1/N)^{p+1}$, while the condition number is quadratic in $N$ and cubic in $p$~\cite{eisentrager2020condition}.

\subsubsection{Variational formulation}
The linear system \eqref{eq:linear_system} can be solved by minimizing a functional that has as a ground state the solution of the system. An example of such a functional is the residual of the linear system
\begin{align}\label{eq:classical_cost}
    J(\bm \phi) &= ||(K+M)\bm \phi - \bm f||^2\\
    &= \bm \phi^\dagger (K+M)^\dagger (K+M) \bm \phi - \bm \phi^\dagger(K+M)^\dagger\bm f\nonumber\\
    &\quad - \bm f^\dagger (K+M)\bm \phi + \bm f^\dagger \bm f.
\end{align}
For simplicity, in the following, we will write $A:=K+M$.
The minimizer $\bm \phi^\star$ of $J$ solves the linear system. Hence, the variational formulation of the finite element method of the Helmholtz problem is written as:
\begin{align}
    \bm \phi^\star = \arg \min_{\bm \phi} &\,\bm \phi^\dagger A^\dagger A \bm \phi - 2\mathrm{Re}(\bm f^\dagger A \bm \phi).
\end{align}

\section{Variational quantum algorithm}
Variational quantum algorithms (VQAs) are hybrid quantum-classical methods that encode the problem solution within a parameterized quantum state, the parameters of which are found by minimizing a problem-specific cost function. The quantum computer evaluates the cost function, while a classical computer updates the parameters accordingly. These algorithms have gained popularity as they are particularly suited for NISQ devices. Parameterized quantum states $\ket{\phi(\bm{\theta})}$ are evolved using parameterized quantum circuits (PQC) $U(\bm \theta)$, that is,
\[
    \ket{\phi(\bm \theta)} = U_\phi(\bm \theta)\ket{0},
\]
where $\bm \theta$ is the set of real parameters. Without loss of generality, the PQC $U(\bm \theta)$ can be expressed as
\begin{equation}\label{eq:PQC}
    U_\phi(\bm \theta) = \prod_l e^{-i\theta_lH_l}V_l,
\end{equation}
where $H_l$ and $V_l$ are Pauli and non-parametrized operators, respectively. The solution $\bm \phi$ and the right hand side $\bm f$ of the Helmholtz problem are therefore embedded in $n$-qubit quantum states (where $2^n = N_\mathrm{dof}$) as follows
\begin{equation}
    \bm \phi \rightarrow r\ket{\phi(\bm \theta)}\quad \mathrm{and}\quad \bm f \rightarrow \ket{f}
\end{equation}
with $r=||\bm \phi||$ and where we assume that $\bm f$ is normalized (if not, the linear system has to be rescaled with respect to $||\bm f||$). The quantum version of the cost function \eqref{eq:classical_cost} writes
\begin{equation}\label{eq:quantum_cost}
    \mathcal{J}(r,\bm \theta)=r^2\bra{\phi(\bm \theta)}A^\dagger A\ket{\phi (\bm \theta)} - 2r\mathrm{Re}(\langle{f}|A\ket{\phi}),
\end{equation}
It can be shown that the amplitude $r$ writes~\cite{sato2021variational}
\begin{align}
    r(\bm \theta) = \frac{\mathrm{Re}(\langle{f}|A\ket{\phi})}{\bra{\phi(\bm \theta)}A^\dagger A \ket{\phi(\bm \theta)}},
\end{align}
which gives
\begin{align}
    \mathcal{J}(\bm \theta) = -\frac{\mathrm{Re}(\langle{f}|A\ket{\phi})^2}{\bra{\phi(\bm \theta)}A^\dagger A\ket{\phi(\bm \theta)}}.
\end{align}
The cost function gradient writes
\begin{align}
        \frac{\partial \mathcal{J}}{\partial \theta_j} = &-2\frac{\mathrm{Re}(\langle{f}|A\ket{\phi})\displaystyle\frac{\partial}{\partial \theta_j}\mathrm{Re}(\langle{f}|A\ket{\phi})}{\bra{\phi(\bm \theta)}A^\dagger A\ket{\phi(\bm \theta)}}\\&+ \frac{\mathrm{Re}(\langle{f}|A\ket{\phi})^2\displaystyle\frac{\partial}{\partial \theta_j}\bra{\phi(\bm \theta)}A^\dagger A\ket{\phi(\bm \theta)}}{\bra{\phi(\bm \theta)}A^\dagger A\ket{\phi(\bm \theta)}^2},
\end{align}
which can be evaluated exactly by differentiating $U_\phi(\bm \theta)$.

\subsection{Efficient decomposition of $A$ and $A^\dagger A$}
For the expectation value $\bra{\phi}A^\dagger A\ket{\phi}$ and the overlap $\bra{f}A\ket{\phi}$ to be computed efficiently, $A$ and $A^\dagger A$ have to be decomposed into a sum of \textit{few} operators that are easy to implement and measure on a quantum computer. For the decomposition to be efficient, the number of terms in the decomposition has to be, at most, polynomial in the number of qubits $n$. Therefore, in this subsection, we compute the decompositions
\begin{align}
    A = \sum_k A_k\quad \mathrm{and}\quad A^\dagger A=\sum_kA'_k.
\end{align}
For regular meshes and Dirichlet and Neumann boundary conditions, one can show that $K$, $M$ and consequentely $A$ and $A^\dagger A$, can be written as block-tridiagonal matrices with constant blocks, leading to the decomposition formulation:
\newpage
\begin{strip}
\begin{align}
\begin{cases}
    A &= 
    \underbrace{I^{\otimes (n-n_B)} \otimes B}_{A_1} + \underbrace{P_B \bigl[ I^{\otimes (n-n_B)} \otimes B \bigr] P_B^{-1}}_{A_2} \underbrace{-P_C \bigl[ I_0^{\otimes (n-n_C)} \otimes C \bigr] P_C^{-1}}_{A_3} + \underbrace{I_0^{\otimes (n-n_{D_L})} \otimes D_L}_{A_4} + \underbrace{I_1^{\otimes (n-n_{D_R})} \otimes D_R}_{A_5},\\
A^\dagger A &= \underbrace{I^{\otimes (n-n_{\tilde B})} \otimes \tilde B}_{A'_1}
  + \underbrace{P_{\tilde B} \bigl[ I^{\otimes (n-n_{\tilde B})} \otimes \tilde B \bigr] P_{\tilde B}^{-1}}_{A'_2} \underbrace{- P_{\tilde C} \bigl[ I_0^{\otimes (n-n_{\tilde C})} \otimes \tilde C \bigr] P_{\tilde C}^{-1}}_{A'_3} + \underbrace{I_0^{\otimes (n-n_{U_L})} \otimes (-I_0\otimes SS^\dagger + U_L)}_{A'_4} \\&+ \underbrace{I_1^{\otimes (n-n_{U_R})} \otimes (-I_1\otimes S^\dagger S + U_R)}_{A'_5},
\end{cases}\nonumber
\end{align}
\end{strip}
\noindent with $n_O := \log_2(\dim(O))$, $P_O := P^{\dim (O)/2}$ and $P$ the circular permutation operator. The derivations together with the definition of $B$, $C$, $\tilde B$, $\tilde C$, $D_L$, $D_R$, $S$, $I_0$, $I_1$, $U_L$ and $U_R$ are given in appendix~A.
Since $S, D_L, D_R\in \mathbb{C}^{p\times p}$, $B,C, U_L, U_R\in \mathbb{C}^{2p\times 2p}$, and $\tilde B,\tilde C\in \mathbb{C}^{4p\times 4p}$, the number of terms of these decompositions are $\mathcal{O}(p^2)$ and do not scale with the number of finite elements $N$.

\subsection{Linear combination of unitaries}
    Measuring $\mathrm{Re}\bra{f}A\ket{\phi}$ requires to encode $A$ in a quantum circuit. As $A$ is non unitary, one has to block-encode it into a larger-space unitary operator. To do so, the first step is to decompose $A$ as a sum of unitary operators:
\begin{align}\label{eq:lcu}
    \begin{cases}
    A &= \sum_j \alpha_j U_j,\\
    A_k &=\sum_{m}\alpha_{km}U_{km},
    \end{cases}
    \Rightarrow A = \sum_k\sum_{m}\alpha_{km}U_{km}.
\end{align}
The product $A\ket{\phi}$ can then be computed
using the following identities:
\begin{align}\label{eq:lcu}
    \begin{cases}
        \hfill A_1 &= I^{\otimes (n-n_B)}\otimes B = \displaystyle\sum_m \alpha_{1m}I^{\otimes (n-n_B)}\otimes U_m^{(B)}\\
        \hfill A_2 &= P_BA_1P_B^{-1} = \displaystyle\sum_m \alpha_{1m}P_B[I^{\otimes (n-n_B)}\otimes U_m^{(B)}]P_B^{-1}\\
        \hfill A_3 &= P_C[I_0^{\otimes (n-n_C)}\otimes C]P_C^{-1} = \displaystyle\sum_m \frac{\alpha_{3m}}{2}I^{\otimes (n-n_C)}\otimes U_m^{(C)}\\&+\displaystyle\sum_m \frac{\alpha_{3m}}{2}R_0^{(n-n_C)}\otimes U_m^{(C)}\\
        \hfill A_4 &= I_0^{\otimes (n-n_{D_L})}\otimes D_L = \displaystyle\sum_m \frac{\alpha_{4m}}{2}I^{\otimes (n-n_{D_L})}\otimes U_m^{(D_L)}\\&+\displaystyle\sum_m \frac{\alpha_{4m}}{2}R_0^{(n-n_{D_L})}\otimes U_m^{(D_L)}\\
        \hfill A_5 &= I_1^{\otimes (n-n_{D_R})}\otimes D_R = \displaystyle\sum_m \frac{\alpha_{5m}}{2}I^{\otimes (n-n_{D_R})}\otimes U_m^{(D_R)}\\&+\displaystyle\sum_m \frac{\alpha_{5m}}{2}R_1^{(n-n_{D_R})}\otimes U_m^{(D_R)},\\
    \end{cases}
\end{align}

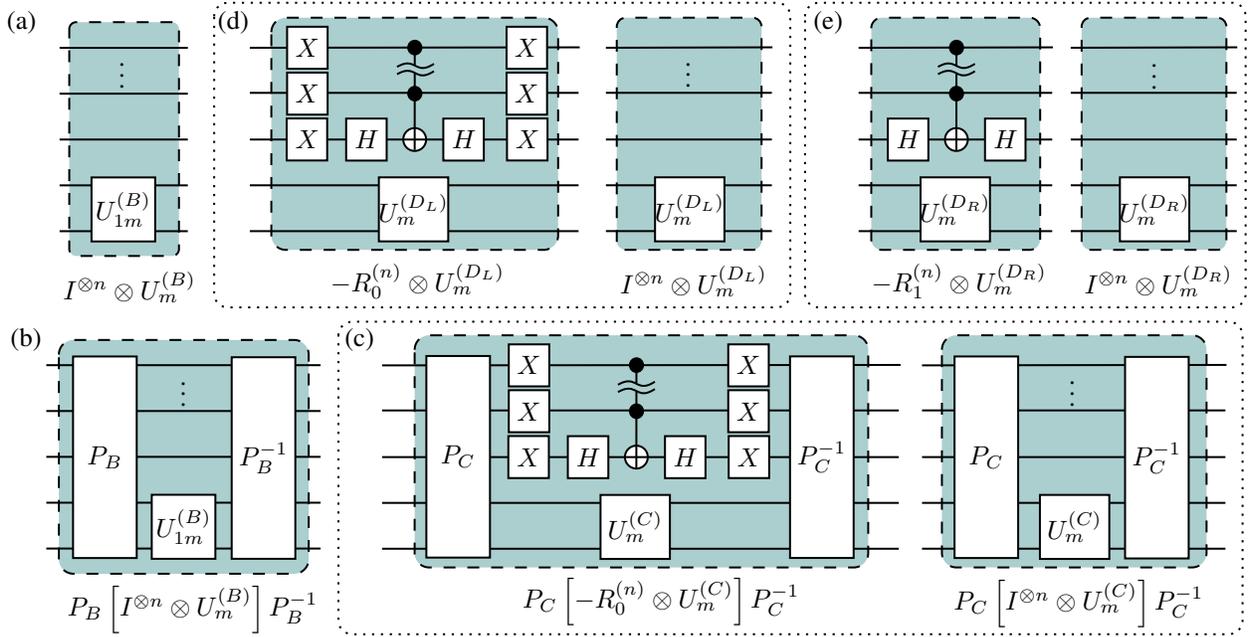
\begin{figure*}[htbp]
    \centering
    \input{LCU.tikz}
    \caption{\textbf{Quantum circuits for the linear combination of unitaries of $A$}. The LCU writes $A_k = \sum_{m}\alpha_{km}U_{km}$ with $k = 1, ..., 5$. The quantum circuits implement the unitary operators required for the LCU of (a) $A_1$, (b) $A_2$, (c) $A_3$, (d) $A_4$, and (e) $A_5$, respectively.}
    \label{fig:lcu}
\end{figure*}

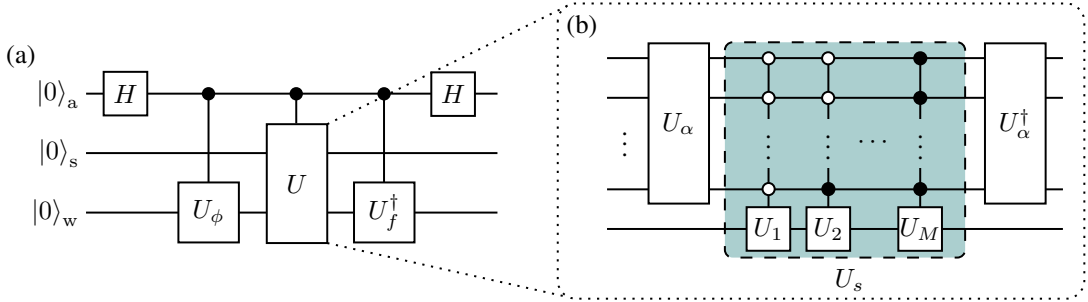
\begin{figure*}[htbp]
    \centering
    \input{block_encoding.tikz}
    \caption{\textbf{Quantum circuit for the measurement of $\mathrm{Re}(\bra f A\ket \phi)$}. 
    (a) Hadamard test. The measuerment of the ancilla qubit (register a) in the computational basis leads to $(1+\mathrm{Re}(\bra{f}A\ket\phi/\eta))/2$.
    (b) Block encoding $U$ of the non unitary matrix $A$. The operator $U_\alpha$ acts on the register s as follows : $U_\alpha \ket{0}_\mathrm{s} = \sum_j (\alpha_j/\eta)^{1/2}\ket{j}_\mathrm{s}$, while $U_s$ acts jointly on registers s and w as follows : $U_s\ket j_\mathrm{s} \ket \phi_\mathrm{w} = \ket j_\mathrm{s} U_j \ket \phi_\mathrm{w}$.}
    \label{fig:block_encoding}
\end{figure*}

The linear combination of unitaries of each $A_k$ is given in \eqref{eq:lcu}, with $R_0:=2I_0-I$, $R_1:=2I_1-I$, where $U_m^{(B)}$, $U_m^{(C)}$, $U_m^{(D_L)}$ and $U_m^{(D_R)}$ can be decomposed in sums of $\mathcal{O}(p^2)$ Pauli strings. The quantum circuits that implement $I^{\otimes n}\otimes U_m^{(B)}$,  $P_B[I^{\otimes n}\otimes U_m^{(B)}]P_B^{-1}$, $I^{\otimes n}\otimes U_m^{(C)}$, $-R_0^{(n)}\otimes U_m^{(C)}$, $-R_0^{(n)}\otimes U_m^{(C)}$, $I^{\otimes n}\otimes U_m^{(D_L)}$, $-R_0^{(n)}\otimes U_m^{(D_L)}$, $I^{\otimes n}\otimes U_m^{(D_R)}$ and $-R_1^{(n)}\otimes U_m^{(D_R)}$ are shown in Figure~\ref{fig:lcu}. Once $A$ is decomposed as a sum of unitary operators, $A\ket{\phi}$ can be computed using a block-encoding of $A$, that is, a unitary operator $U$ such that
\begin{align}
    \bra{0}U\ket{0}\ket{\phi} = \frac{A}{\eta}\ket{\phi},\quad U=
    \begin{pmatrix}
        A/\eta & *\\
        * & *
    \end{pmatrix},
\end{align}
with $\eta:=\sum_{km}|\alpha_{km}|$. Formally, the unitary $U$ can be expressed as
\begin{align}
    U = (U_\alpha^\dagger \otimes I^{\otimes n}) U_s (U_\alpha \otimes I^{\otimes n}),
\end{align}
with $U_\alpha \ket{0} = \sum_j (\alpha_j/\eta)^{1/2}\ket{j}$ and $U_s\ket j \ket \phi = \ket j U_j \ket \phi$. Once $U$ is designed, $\mathrm{Re}(\bra fA\ket \phi)$ can be computed using the Hadamard-test-like circuit shown in Figure~\ref{fig:block_encoding}(a). The quantum circuit for the block-encoding $U$ of $A$ is shown in Figure~\ref{fig:block_encoding}(b). The quantum circuit for the measurement of $\bra{\phi}A^\dagger A\ket \phi$ is analogous to the one of Figure~\ref{fig:block_encoding}, but with $U_\phi^\dagger$ instead of $U_f^\dagger$ and $U'$ instead of $U$, where $U'$ is the block encoding of $A^\dagger A$. 


\section{Results and discussion}
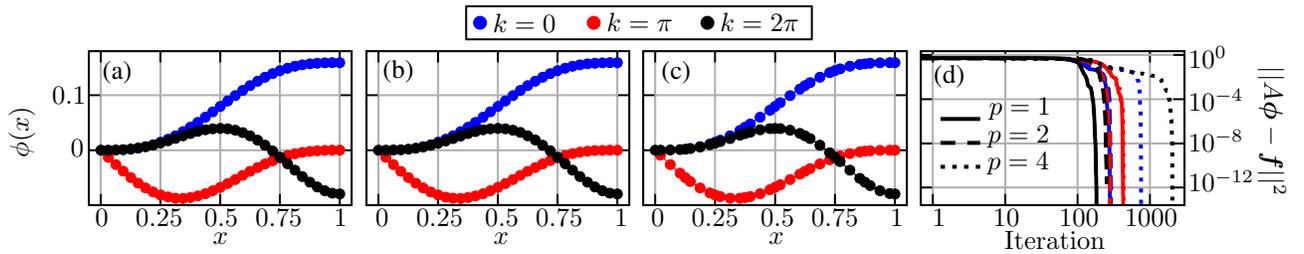
\begin{figure*}[htbp]
    \centering
    \input{result.tikz}
    \caption{\textbf{Variational quantum algorithm solution for the Helmholtz problem \eqref{eq:linear_system}.} The PQC architecture is the hardware efficient ansatz (HEA)~\cite{kandala2017hardware}, with $7$ layers of $R_Y(\theta_j)$ rotation gates with linear sequences of CNOT entanglement gates. The order of the finite elements is (a) $p=1$, (b) $p=2$, (c) $p=4$. The number of degrees of freedom is kept constant. (d) Square norm of the residual.}
    \label{fig:results}
\end{figure*}

We solved the Helmholtz problem \eqref{eq:hermitian_problem} for wave numbers in $k\in\{0, \pi, 2\pi\}$, and with first, second and fourth order finite elements. The results are shown in Figure~\ref{fig:results}. 
The PQC architecture has been fixed to the hardware efficient ansatz (HEA)~\cite{kandala2017hardware}, with only $R_Y$ parameterized gates to constraint the solution to the real plane, and with linear sequences of CNOT entanglement gates. The cost function has been minimized using the Broyden-Fletcher-Goldfarb-Shanno (BFGS) algorithm, initialized at $\bm \theta_{0} = \bm 0$.

\subsection{Ansatz expressiveness}
The expressiveness of $U_\phi$ over a group $G$ is its ability to represent the operators spanned by $G$. Haar-random states are quantum states sampled uniformly, in the Haar measure from the group $G$. Hence, a common tool for evaluating the expressiveness of a PQC is the Kullback-Leibler divergence $D_\mathrm{KL}$ between the distribution of the states sampled by the PQC $\rho_\phi$ and the Haar distribution $\rho_\mathrm{Haar}$~\cite{sim2019expressibility}. This measure quantifies the discrepancy of the distribution of states sampled from $U_\phi$ and from the Haar measure, respectively. 
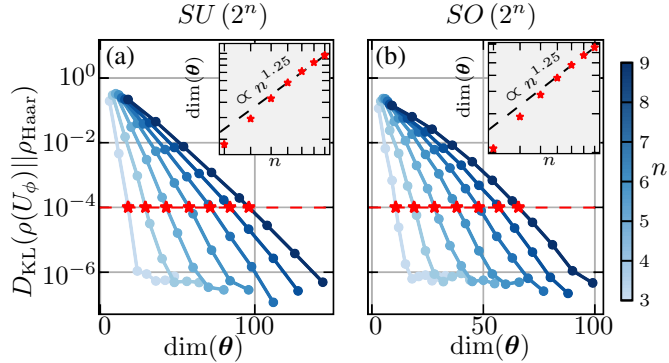
\begin{figure}[htbp]
    \centering
    \input{kldiv.tikz}
    \caption{\textbf{Kullback-Leibler divergence between states sampled from the HEA and Haar random states.} (a) HEA with $R_Y$ and $R_Z$ rotation gates compared to Haar random states over $SU(2^n)$. (b) HEA with $R_Y$ rotation gates compared to Haar random states over $SO(2^n)$.}
    \label{fig:kldiv}
\end{figure}
In Figure~\ref{fig:kldiv}, we show this KL divergence for the HEA, with increasing the number of parameterized quantum gates, for different size of PQCs. We observe that the number of parameters increases polynomially with the number of qubits for maintaining the same expressiveness. 

\subsection{Algorithm scaling}
An iteration of our algorithm consists in evaluating the cost function gradient and updating the set of parameters. The computational cost of the gradient evaluation is $\mathcal{O}(\dim(\bm \theta)d/\varepsilon^2)$, with $d$ the depth of the quantum circuits and $\varepsilon$ the absolute precision. 
Only two different quantum circuits are required for the gradient evaluation: the Hadamard test for $\mathrm{Re}(\bra{f}A\ket\phi)$ and the Hadamard test for $\bra{\phi}A^\dagger A\ket \phi$.
\paragraph*{Depth of the Hadamard test for $\mathrm{Re}(\bra{f}A\ket\phi)$}
The depth of the Hadamard test circuit (see Figure~\ref{fig:block_encoding}) depends  on the depth of $U_\phi$, $U_f$, and $U$ (and, by extension, $U_\alpha$ and $U_s$):
\begin{itemize}
    \item[i.] We make the common assumption that 
    there exists an efficient quantum circuit $U_f$ of depth $\mathcal{O}(n)$~\cite{sato2021variational};
    \item[ii.] The depth of $U_\phi$ depends on the number and size of the layers. The number of layers $L$ is proportional to $\dim(\bm\theta)/n$. Since we observed that $\dim(\theta)=\mathcal{O}(n^{1.25})$ we consider that $ L = \mathcal{O}(n^{1/4})$. Each linear entanglement layer is of depth $\mathcal{O}(n)$, so the total depth of $U_\phi$ is $\mathcal{O}(n^{1.25})$;
    \item[iii.] As $U_\alpha$ acts on a $\mathcal{O}(\log p)$-qubit register, we expect that its depth is $\mathcal{O}(\mathrm{poly}\log p)$;
    \item[iv.] The depth and number of circuits for each the unitary decomposition of each $A_k$ is shown in Table~\ref{tab:lcu}. The depth of $U_s$ directly follows and scales as $\mathcal{O}(p^3n^2)$.
\end{itemize}
These depth scalings are also valid for the Hadamard test that measures $\bra{\phi}A^\dagger A\ket{\phi}$. Hence the computational cost of a gradient evaluation of the cost function is $\mathcal{O}(p^3n^2/\varepsilon^2)$. Recall that $n = \log_2N_\mathrm{dof}$ and $N_\mathrm{dof}\sim Np$, it leads to an iteration cost of $\mathcal{O}(p^3\mathrm{polylog}(Np)/\varepsilon^2)$.
\begin{table}[h!]
    \centering
    \begin{tabular}{|c|c|c|}
        \hline
         &  \# circuits &  depth $d$ \\\hline
        $A_1,\, A'_1$ & $\mathcal{O}(p^2)$ & $\mathcal{O}(1)$\\\hline
        $A_2,\, A'_2$ & $\mathcal{O}(p^2)$ & $\mathcal{O}(pn^2)$ \\\hline
        $A_3,\, A'_3$ & $\mathcal{O}(p^2)$ & $\mathcal{O}(pn^2 + n) + \mathcal{O}(pn^2)$ \\\hline
        $A_4,\, A'_4$ & $\mathcal{O}(p^2)$ & $\mathcal{O}(n) + \mathcal{O}(1)$\\\hline
        $A_5,\, A'_5$ & $\mathcal{O}(p^2)$ & $\mathcal{O}(n) + \mathcal{O}(1)$\\\hline
        \end{tabular}
    \caption{\textbf{Quantum circuit depth and number for the linear combination of unitaries of $A$.} }
    \label{tab:lcu}
\end{table}

\section{Conclusion}
We proposed a variational quantum algorithm for solving the Helmholtz problem with high order finite elements on a regular mesh. The computational cost of an iteration is cubic with respect to the finite element order $p$, and logarithmic in the number of elements $N$. We solved a one-dimensional Helmholtz problem with Dirichlet and Neumann boundary conditions, for various wave numbers. We also discussed the scaling of the KL divergence of the ansatz with respect to Haar distributed states, and observed that the number of parameters scales polynomially with the number of qubits, leading to the resolution of a logarithmic-sized non-convex optimization problem with respect to the number of degrees of freedom. Future works include the extension to other boundary conditions (e.g., open systems), and to space-dependent wavenumbers (e.g., piece-wise constant).
\begin{strip}
\section*{Appendix}
\subsection{Mathematical developments}\label{app:maths}
\subsubsection*{Decomposition of $A$} We have
\begin{align}
A &=
\begin{pmatrix}
D + D_L   & S                  &     & \\
S^\dagger & D                  & \ddots    &  \\
    & \ddots         & \ddots    & S \\
        &              & S^\dagger & D + D_R
\end{pmatrix}
=A_{\mathrm{int}} + A_\mathrm{bc}=
\begin{pmatrix}
D    & S                  &     & \\
S^\dagger & D                  & \ddots    &  \\
    & \ddots         & \ddots    & S \\
        &              & S^\dagger & D
\end{pmatrix}
+
\begin{pmatrix}
D_L    &                  &     & \\
  & 0                  &    &  \\
    &          & \ddots    & \\
        &              &  & D_R
\end{pmatrix}
\end{align}

\begin{align}
    \Leftrightarrow A 
    &= I^{\otimes (n-n_B)} \otimes B
  +
  P_B \bigl[ I^{\otimes (n-n_B)} \otimes B \bigr] P_B^{-1}
  -P_C \bigl[ I_0^{\otimes (n-n_C)} \otimes C \bigr] P_C^{-1}
  +
  I_0^{\otimes (n-n_{D_L})} \otimes D_L  +
  I_1^{\otimes (n-n_{D_R})} \otimes D_R,
\end{align}

with $I_0 = \ket{0}\bra{0}$, $I_1=\ket{1}\bra{1}$, $I$ the identity, and  

\begin{align}
    B = 
    \begin{pmatrix}
        D/2&S\\
        S^\dagger&D/2
    \end{pmatrix}
    \quad \mathrm{and} \quad 
    C = 
    \begin{pmatrix}
        0&S\\
        S^\dagger&0
    \end{pmatrix}.
\end{align}
\subsubsection*{Decomposition of $A^\dagger A$}
We have
\begin{align*}
A^\dagger A
&= (A_{\mathrm{int}}^\dagger + A_{\mathrm{bc}}^\dagger)\,(A_{\mathrm{int}} + A_{\mathrm{bc}}) = A_{\mathrm{int}}^\dagger A_{\mathrm{int}}
 + A_{\mathrm{int}}^\dagger A_{\mathrm{bc}}
 + A_{\mathrm{bc}}^\dagger A_{\mathrm{int}}
 + A_{\mathrm{bc}}^\dagger A_{\mathrm{bc}}.
\end{align*}
It can be shown that by defining 
\begin{align}
    U = 
      \begin{pmatrix}
            S^\dagger S + D^\dagger D + SS^\dagger & D^\dagger S+ SD\\
            S^\dagger D+ D^\dagger S^\dagger & S^\dagger S + D^\dagger D + SS^\dagger
      \end{pmatrix}\quad \mathrm{and}\quad
    V =
      \begin{pmatrix}
            S^2 & 0\\
            D^\dagger S+ SD & S^2
      \end{pmatrix},
\end{align}
we can write 
\begin{align}
      A_{\mathrm{int}}^\dagger A_{\mathrm{int}} =
      \begin{pmatrix}
        U         & V                 &     &  \\
        V^\dagger & U                 & \ddots    &  \\
            & \ddots            & \ddots    & V \\
                 &             & V^\dagger & U
      \end{pmatrix}
      +
      \begin{pmatrix}
        -SS^\dagger  &                  &     &  \\
                    & 0                 &     & \\
               &             & \ddots    &  \\
                    &             &          & -S^\dagger S
      \end{pmatrix}\nonumber
&= 
  I^{\otimes (n-n_{\tilde B})} \otimes \tilde B
  \;+\;
  P_{\tilde B} \bigl[ I^{\otimes (n-n_{\tilde B})} \otimes \tilde B \bigr] P_{\tilde B}^{-1}\\[-2.8em]
  &-
  P_{\tilde C} \bigl[ I_0^{\otimes (n-n_{\tilde C})} \otimes \tilde C \bigr] P_{\tilde C}^{-1}\nonumber\\[0.2em]
  &+
  I_0^{\otimes n} \otimes (-SS^\dagger)
  \;+\;
  I_1^{\otimes n} \otimes (-S^\dagger S),
\end{align}

and
\begin{equation}
      A_\mathrm{int}^\dagger A_\mathrm{bc} + A_\mathrm{bc}^\dagger A_\mathrm{int} + A_\mathrm{bc}^\dagger A_\mathrm{bc} = I_0^{\otimes n}\otimes U_L + I_1^{\otimes n}\otimes U_R,
\end{equation}
with
\begin{align}
    \begin{cases}
            \tilde B = 
    \begin{pmatrix}
        U/2&V\\
        V^\dagger&U/2
    \end{pmatrix},\quad
    U_L = 
    \begin{pmatrix}
        D^\dagger D_L + D_L^\dagger D + D_L^\dagger D_L&D_L^\dagger S\\
        S^\dagger D_L&0
    \end{pmatrix}\\
    \tilde C = 
    \begin{pmatrix}
        0&V\\
        V^\dagger&0
    \end{pmatrix},\quad 
    U_R=
    \begin{pmatrix}
        0&S D_R\\
        D_R^\dagger S^\dagger &D^\dagger D_R + D_R^\dagger D + D_R^\dagger D_R
    \end{pmatrix}.
    \end{cases}
\end{align}
The total decomposition of $A^\dagger A$ writes
\begin{align}
    A^\dagger A 
&= 
    I^{\otimes n} \otimes \tilde B
  \;+\;
  P_{\tilde B} \bigl[ I^{\otimes n} \otimes \tilde B \bigr] P_{\tilde B}^{-1}
    - P_{\tilde C} \bigl[ I_0^{\otimes n} \otimes \tilde C \bigr] P_{\tilde C}^{-1}
  \;+\;
  I_0^{\otimes n} \otimes (-I_0\otimes SS^\dagger + U_L)
  \;+\;
I_1^{\otimes n} \otimes (-I_1\otimes S^\dagger S + U_R),
\end{align}
\end{strip}

\FloatBarrier
\bibliographystyle{ieeetr}
\bibliography{ref}

\end{document}

%% file: LCU.tikz
\tikzset{every picture/.style={line width=0.75pt}} 

\begin{tikzpicture}[x=0.75pt,y=0.75pt,yscale=-1,xscale=1]

\draw  [color={rgb, 255:red, 0; green, 0; blue, 0 }  ,draw opacity=1 ][fill={rgb, 255:red, 171; green, 207; blue, 207 }  ,fill opacity=1 ][dash pattern={on 4.5pt off 4.5pt}] (133.83,69.57) .. controls (133.83,65.54) and (137.11,62.26) .. (141.15,62.26) -- (271.15,62.26) .. controls (275.19,62.26) and (278.47,65.54) .. (278.47,69.57) -- (278.47,172.13) .. controls (278.47,176.17) and (275.19,179.44) .. (271.15,179.44) -- (141.15,179.44) .. controls (137.11,179.44) and (133.83,176.17) .. (133.83,172.13) -- cycle ;
\draw  [fill={rgb, 255:red, 255; green, 255; blue, 255 }  ,fill opacity=1 ] (200.36,123.74) .. controls (200.36,120.55) and (202.94,117.97) .. (206.12,117.97) .. controls (209.31,117.97) and (211.89,120.55) .. (211.89,123.74) .. controls (211.89,126.92) and (209.31,129.51) .. (206.12,129.51) .. controls (202.94,129.51) and (200.36,126.92) .. (200.36,123.74) -- cycle ;
\draw  [color={rgb, 255:red, 0; green, 0; blue, 0 }  ,draw opacity=1 ][fill={rgb, 255:red, 171; green, 207; blue, 207 }  ,fill opacity=1 ][dash pattern={on 4.5pt off 4.5pt}] (26.67,232.28) .. controls (26.67,228.21) and (29.96,224.91) .. (34.03,224.91) -- (145.3,224.91) .. controls (149.37,224.91) and (152.67,228.21) .. (152.67,232.28) -- (152.67,335.55) .. controls (152.67,339.62) and (149.37,342.92) .. (145.3,342.92) -- (34.03,342.92) .. controls (29.96,342.92) and (26.67,339.62) .. (26.67,335.55) -- cycle ;
\draw  [color={rgb, 255:red, 0; green, 0; blue, 0 }  ,draw opacity=1 ][fill={rgb, 255:red, 171; green, 207; blue, 207 }  ,fill opacity=1 ][dash pattern={on 4.5pt off 4.5pt}] (31.92,67.97) .. controls (31.92,66.06) and (33.46,64.51) .. (35.37,64.51) -- (83.79,64.51) .. controls (85.7,64.51) and (87.24,66.06) .. (87.24,67.97) -- (87.24,179.06) .. controls (87.24,180.97) and (85.7,182.52) .. (83.79,182.52) -- (35.37,182.52) .. controls (33.46,182.52) and (31.92,180.97) .. (31.92,179.06) -- cycle ;
\draw    (27.03,77.46) -- (91.7,77.46) ;
\draw    (27.03,100.6) -- (91.7,100.6) ;
\draw    (27.03,123.74) -- (91.7,123.74) ;
\draw    (27.03,146.88) -- (91.7,146.88) ;
\draw    (27.03,170.01) -- (91.7,170.01) ;
\draw  [fill={rgb, 255:red, 255; green, 255; blue, 255 }  ,fill opacity=1 ] (43.14,142.87) -- (75.24,142.87) -- (75.24,175.19) -- (43.14,175.19) -- cycle ;
\draw    (20.33,237.64) -- (158.67,237.64) ;
\draw    (20.33,260.78) -- (158.67,260.78) ;
\draw    (20.33,283.91) -- (158.67,283.91) ;
\draw    (20.33,307.05) -- (158.67,307.05) ;
\draw    (20.33,330.19) -- (158.67,330.19) ;
\draw  [fill={rgb, 255:red, 255; green, 255; blue, 255 }  ,fill opacity=1 ] (73.51,303.05) -- (105.6,303.05) -- (105.6,335.37) -- (73.51,335.37) -- cycle ;
\draw  [fill={rgb, 255:red, 255; green, 255; blue, 255 }  ,fill opacity=1 ] (33.83,233.25) -- (65.93,233.25) -- (65.93,334.98) -- (33.83,334.98) -- cycle ;
\draw  [fill={rgb, 255:red, 255; green, 255; blue, 255 }  ,fill opacity=1 ] (113.83,233.63) -- (145.93,233.63) -- (145.93,335.37) -- (113.83,335.37) -- cycle ;
\draw    (122.97,77.44) -- (289.36,77.22) ;
\draw    (122.97,100.19) -- (209.7,100.19) -- (230.14,100.19) -- (230.32,100.19) -- (288.78,100.19) ;
\draw    (122.97,123.74) -- (288.78,123.74) ;
\draw    (122.97,146.85) -- (288.78,146.85) ;
\draw    (122.97,169.99) -- (288.78,169.99) ;
\draw  [fill={rgb, 255:red, 255; green, 255; blue, 255 }  ,fill opacity=1 ] (188.05,142.83) -- (223.05,142.83) -- (223.05,175.15) -- (188.05,175.15) -- cycle ;
\draw  [fill={rgb, 255:red, 255; green, 255; blue, 255 }  ,fill opacity=1 ] (141.56,113.11) -- (161.99,113.11) -- (161.99,134.32) -- (141.56,134.32) -- cycle ;
\draw  [fill={rgb, 255:red, 255; green, 255; blue, 255 }  ,fill opacity=1 ] (141.78,89.97) -- (162.21,89.97) -- (162.21,111.18) -- (141.78,111.18) -- cycle ;
\draw  [fill={rgb, 255:red, 255; green, 255; blue, 255 }  ,fill opacity=1 ] (141.78,66.83) -- (162.21,66.83) -- (162.21,88.04) -- (141.78,88.04) -- cycle ;
\draw  [fill={rgb, 255:red, 255; green, 255; blue, 255 }  ,fill opacity=1 ] (252.68,113.11) -- (273.11,113.11) -- (273.11,134.32) -- (252.68,134.32) -- cycle ;
\draw  [fill={rgb, 255:red, 255; green, 255; blue, 255 }  ,fill opacity=1 ] (252.46,89.97) -- (272.89,89.97) -- (272.89,111.18) -- (252.46,111.18) -- cycle ;
\draw  [fill={rgb, 255:red, 255; green, 255; blue, 255 }  ,fill opacity=1 ] (252.46,66.83) -- (272.89,66.83) -- (272.89,88.04) -- (252.46,88.04) -- cycle ;
\draw  [fill={rgb, 255:red, 0; green, 0; blue, 0 }  ,fill opacity=1 ] (202.87,100.57) .. controls (202.87,98.67) and (204.35,97.13) .. (206.18,97.13) .. controls (208.02,97.13) and (209.5,98.67) .. (209.5,100.57) .. controls (209.5,102.47) and (208.02,104.02) .. (206.18,104.02) .. controls (204.35,104.02) and (202.87,102.47) .. (202.87,100.57) -- cycle ;
\draw  [fill={rgb, 255:red, 0; green, 0; blue, 0 }  ,fill opacity=1 ] (202.87,77.33) .. controls (202.87,75.42) and (204.35,73.88) .. (206.18,73.88) .. controls (208.02,73.88) and (209.5,75.42) .. (209.5,77.33) .. controls (209.5,79.23) and (208.02,80.77) .. (206.18,80.77) .. controls (204.35,80.77) and (202.87,79.23) .. (202.87,77.33) -- cycle ;
\draw    (206.16,77.33) -- (206.12,129.06) ;
\draw  [fill={rgb, 255:red, 255; green, 255; blue, 255 }  ,fill opacity=1 ] (171.56,113.16) -- (191.99,113.16) -- (191.99,134.37) -- (171.56,134.37) -- cycle ;
\draw  [fill={rgb, 255:red, 255; green, 255; blue, 255 }  ,fill opacity=1 ] (220.56,113.16) -- (240.99,113.16) -- (240.99,134.37) -- (220.56,134.37) -- cycle ;
\draw  [color={rgb, 255:red, 0; green, 0; blue, 0 }  ,draw opacity=1 ][fill={rgb, 255:red, 171; green, 207; blue, 207 }  ,fill opacity=1 ][dash pattern={on 4.5pt off 4.5pt}] (308.04,66.81) .. controls (308.04,64.29) and (310.08,62.25) .. (312.6,62.25) -- (376.5,62.25) .. controls (379.02,62.25) and (381.06,64.29) .. (381.06,66.81) -- (381.06,174.87) .. controls (381.06,177.39) and (379.02,179.43) .. (376.5,179.43) -- (312.6,179.43) .. controls (310.08,179.43) and (308.04,177.39) .. (308.04,174.87) -- cycle ;
\draw    (302.56,77.43) -- (386.56,77.21) ;
\draw    (302.56,100.17) -- (346.34,100.17) -- (356.66,100.17) -- (356.75,100.17) -- (386.26,100.17) ;
\draw    (302.56,123.73) -- (386.26,123.73) ;
\draw    (302.56,146.84) -- (386.26,146.84) ;
\draw    (302.56,169.97) -- (386.26,169.97) ;
\draw  [fill={rgb, 255:red, 255; green, 255; blue, 255 }  ,fill opacity=1 ] (327.35,142.82) -- (362.35,142.82) -- (362.35,175.14) -- (327.35,175.14) -- cycle ;
\draw  [color={rgb, 255:red, 0; green, 0; blue, 0 }  ,draw opacity=1 ][dash pattern={on 0.84pt off 2.51pt}] (105,61.57) .. controls (105,57.79) and (108.07,54.72) .. (111.85,54.72) -- (388.71,54.72) .. controls (392.49,54.72) and (395.56,57.79) .. (395.56,61.57) -- (395.56,201.62) .. controls (395.56,205.41) and (392.49,208.47) .. (388.71,208.47) -- (111.85,208.47) .. controls (108.07,208.47) and (105,205.41) .. (105,201.62) -- cycle ;
\draw  [color={rgb, 255:red, 0; green, 0; blue, 0 }  ,draw opacity=1 ][fill={rgb, 255:red, 171; green, 207; blue, 207 }  ,fill opacity=1 ][dash pattern={on 4.5pt off 4.5pt}] (206.1,229.87) .. controls (206.1,225.84) and (209.38,222.56) .. (213.42,222.56) -- (427.55,222.56) .. controls (431.58,222.56) and (434.86,225.84) .. (434.86,229.87) -- (434.86,332.43) .. controls (434.86,336.47) and (431.58,339.74) .. (427.55,339.74) -- (213.42,339.74) .. controls (209.38,339.74) and (206.1,336.47) .. (206.1,332.43) -- cycle ;
\draw  [fill={rgb, 255:red, 255; green, 255; blue, 255 }  ,fill opacity=1 ] (312.16,284.04) .. controls (312.16,280.85) and (314.74,278.27) .. (317.93,278.27) .. controls (321.11,278.27) and (323.69,280.85) .. (323.69,284.04) .. controls (323.69,287.22) and (321.11,289.81) .. (317.93,289.81) .. controls (314.74,289.81) and (312.16,287.22) .. (312.16,284.04) -- cycle ;
\draw    (189.68,237.74) -- (451.19,237.52) ;
\draw    (189.68,260.49) -- (325.99,260.49) -- (358.11,260.49) -- (358.4,260.49) -- (450.28,260.49) ;
\draw    (189.68,284.04) -- (450.28,284.04) ;
\draw    (189.68,307.15) -- (450.28,307.15) ;
\draw    (189.68,330.29) -- (450.28,330.29) ;
\draw  [fill={rgb, 255:red, 255; green, 255; blue, 255 }  ,fill opacity=1 ] (299.85,303.13) -- (334.85,303.13) -- (334.85,335.45) -- (299.85,335.45) -- cycle ;
\draw  [fill={rgb, 255:red, 255; green, 255; blue, 255 }  ,fill opacity=1 ] (253.36,273.41) -- (273.79,273.41) -- (273.79,294.62) -- (253.36,294.62) -- cycle ;
\draw  [fill={rgb, 255:red, 255; green, 255; blue, 255 }  ,fill opacity=1 ] (253.58,250.27) -- (274.01,250.27) -- (274.01,271.48) -- (253.58,271.48) -- cycle ;
\draw  [fill={rgb, 255:red, 255; green, 255; blue, 255 }  ,fill opacity=1 ] (253.58,227.13) -- (274.01,227.13) -- (274.01,248.34) -- (253.58,248.34) -- cycle ;
\draw  [fill={rgb, 255:red, 255; green, 255; blue, 255 }  ,fill opacity=1 ] (364.48,273.41) -- (384.91,273.41) -- (384.91,294.62) -- (364.48,294.62) -- cycle ;
\draw  [fill={rgb, 255:red, 255; green, 255; blue, 255 }  ,fill opacity=1 ] (364.26,250.27) -- (384.69,250.27) -- (384.69,271.48) -- (364.26,271.48) -- cycle ;
\draw  [fill={rgb, 255:red, 255; green, 255; blue, 255 }  ,fill opacity=1 ] (364.26,227.13) -- (384.69,227.13) -- (384.69,248.34) -- (364.26,248.34) -- cycle ;
\draw  [fill={rgb, 255:red, 0; green, 0; blue, 0 }  ,fill opacity=1 ] (314.67,260.87) .. controls (314.67,258.97) and (316.15,257.43) .. (317.99,257.43) .. controls (319.82,257.43) and (321.3,258.97) .. (321.3,260.87) .. controls (321.3,262.77) and (319.82,264.32) .. (317.99,264.32) .. controls (316.15,264.32) and (314.67,262.77) .. (314.67,260.87) -- cycle ;
\draw  [fill={rgb, 255:red, 0; green, 0; blue, 0 }  ,fill opacity=1 ] (314.67,237.63) .. controls (314.67,235.72) and (316.15,234.18) .. (317.99,234.18) .. controls (319.82,234.18) and (321.3,235.72) .. (321.3,237.63) .. controls (321.3,239.53) and (319.82,241.07) .. (317.99,241.07) .. controls (316.15,241.07) and (314.67,239.53) .. (314.67,237.63) -- cycle ;
\draw    (317.96,237.63) -- (317.92,289.36) ;
\draw  [fill={rgb, 255:red, 255; green, 255; blue, 255 }  ,fill opacity=1 ] (283.36,273.46) -- (303.79,273.46) -- (303.79,294.67) -- (283.36,294.67) -- cycle ;
\draw  [fill={rgb, 255:red, 255; green, 255; blue, 255 }  ,fill opacity=1 ] (332.36,273.46) -- (352.79,273.46) -- (352.79,294.67) -- (332.36,294.67) -- cycle ;
\draw  [color={rgb, 255:red, 0; green, 0; blue, 0 }  ,draw opacity=1 ][fill={rgb, 255:red, 171; green, 207; blue, 207 }  ,fill opacity=1 ][dash pattern={on 4.5pt off 4.5pt}] (472.02,229.86) .. controls (472.02,225.82) and (475.29,222.55) .. (479.33,222.55) -- (598.18,222.55) .. controls (602.22,222.55) and (605.49,225.82) .. (605.49,229.86) -- (605.49,332.42) .. controls (605.49,336.46) and (602.22,339.73) .. (598.18,339.73) -- (479.33,339.73) .. controls (475.29,339.73) and (472.02,336.46) .. (472.02,332.42) -- cycle ;
\draw    (461.99,237.73) -- (615.55,237.51) ;
\draw    (461.99,260.47) -- (542.03,260.47) -- (560.89,260.47) -- (561.06,260.47) -- (615.01,260.47) ;
\draw    (461.99,284.03) -- (615.01,284.03) ;
\draw    (461.99,307.14) -- (615.01,307.14) ;
\draw    (461.99,330.27) -- (615.01,330.27) ;
\draw  [fill={rgb, 255:red, 255; green, 255; blue, 255 }  ,fill opacity=1 ] (521.45,303.12) -- (556.45,303.12) -- (556.45,335.44) -- (521.45,335.44) -- cycle ;
\draw  [fill={rgb, 255:red, 255; green, 255; blue, 255 }  ,fill opacity=1 ] (212.01,233.01) -- (244.1,233.01) -- (244.1,334.75) -- (212.01,334.75) -- cycle ;
\draw  [fill={rgb, 255:red, 255; green, 255; blue, 255 }  ,fill opacity=1 ] (395.42,233.2) -- (427.51,233.2) -- (427.51,334.94) -- (395.42,334.94) -- cycle ;
\draw  [fill={rgb, 255:red, 255; green, 255; blue, 255 }  ,fill opacity=1 ] (478.39,233.24) -- (510.48,233.24) -- (510.48,334.97) -- (478.39,334.97) -- cycle ;
\draw  [fill={rgb, 255:red, 255; green, 255; blue, 255 }  ,fill opacity=1 ] (564.39,233.46) -- (596.48,233.46) -- (596.48,335.19) -- (564.39,335.19) -- cycle ;
\draw  [color={rgb, 255:red, 0; green, 0; blue, 0 }  ,draw opacity=1 ][dash pattern={on 0.84pt off 2.51pt}] (167.67,222.72) .. controls (167.67,218.83) and (170.81,215.69) .. (174.69,215.69) -- (616.63,215.69) .. controls (620.51,215.69) and (623.66,218.83) .. (623.66,222.72) -- (623.66,366.41) .. controls (623.66,370.29) and (620.51,373.44) .. (616.63,373.44) -- (174.69,373.44) .. controls (170.81,373.44) and (167.67,370.29) .. (167.67,366.41) -- cycle ;
\draw  [color={rgb, 255:red, 0; green, 0; blue, 0 }  ,draw opacity=1 ][fill={rgb, 255:red, 171; green, 207; blue, 207 }  ,fill opacity=1 ][dash pattern={on 4.5pt off 4.5pt}] (436.4,67.58) .. controls (436.4,64.64) and (438.78,62.26) .. (441.72,62.26) -- (516.27,62.26) .. controls (519.2,62.26) and (521.58,64.64) .. (521.58,67.58) -- (521.58,174.12) .. controls (521.58,177.06) and (519.2,179.44) .. (516.27,179.44) -- (441.72,179.44) .. controls (438.78,179.44) and (436.4,177.06) .. (436.4,174.12) -- cycle ;
\draw  [fill={rgb, 255:red, 255; green, 255; blue, 255 }  ,fill opacity=1 ] (473.56,123.74) .. controls (473.56,120.55) and (476.14,117.97) .. (479.32,117.97) .. controls (482.51,117.97) and (485.09,120.55) .. (485.09,123.74) .. controls (485.09,126.92) and (482.51,129.51) .. (479.32,129.51) .. controls (476.14,129.51) and (473.56,126.92) .. (473.56,123.74) -- cycle ;
\draw    (430,77.44) -- (528,77.22) ;
\draw    (430,100.19) -- (481.08,100.19) -- (493.12,100.19) -- (493.23,100.19) -- (527.66,100.19) ;
\draw    (430,123.74) -- (527.66,123.74) ;
\draw    (430,146.85) -- (527.66,146.85) ;
\draw    (430,169.99) -- (527.66,169.99) ;
\draw  [fill={rgb, 255:red, 255; green, 255; blue, 255 }  ,fill opacity=1 ] (461.25,142.83) -- (496.25,142.83) -- (496.25,175.15) -- (461.25,175.15) -- cycle ;
\draw  [fill={rgb, 255:red, 0; green, 0; blue, 0 }  ,fill opacity=1 ] (476.07,100.57) .. controls (476.07,98.67) and (477.55,97.13) .. (479.38,97.13) .. controls (481.22,97.13) and (482.7,98.67) .. (482.7,100.57) .. controls (482.7,102.47) and (481.22,104.02) .. (479.38,104.02) .. controls (477.55,104.02) and (476.07,102.47) .. (476.07,100.57) -- cycle ;
\draw  [fill={rgb, 255:red, 0; green, 0; blue, 0 }  ,fill opacity=1 ] (476.07,77.33) .. controls (476.07,75.42) and (477.55,73.88) .. (479.38,73.88) .. controls (481.22,73.88) and (482.7,75.42) .. (482.7,77.33) .. controls (482.7,79.23) and (481.22,80.77) .. (479.38,80.77) .. controls (477.55,80.77) and (476.07,79.23) .. (476.07,77.33) -- cycle ;
\draw    (479.36,77.33) -- (479.32,129.06) ;
\draw  [fill={rgb, 255:red, 255; green, 255; blue, 255 }  ,fill opacity=1 ] (444.76,113.16) -- (465.19,113.16) -- (465.19,134.37) -- (444.76,134.37) -- cycle ;
\draw  [fill={rgb, 255:red, 255; green, 255; blue, 255 }  ,fill opacity=1 ] (493.76,113.16) -- (514.19,113.16) -- (514.19,134.37) -- (493.76,134.37) -- cycle ;
\draw  [color={rgb, 255:red, 0; green, 0; blue, 0 }  ,draw opacity=1 ][fill={rgb, 255:red, 171; green, 207; blue, 207 }  ,fill opacity=1 ][dash pattern={on 4.5pt off 4.5pt}] (542.49,66.81) .. controls (542.49,64.29) and (544.53,62.25) .. (547.04,62.25) -- (610.94,62.25) .. controls (613.46,62.25) and (615.5,64.29) .. (615.5,66.81) -- (615.5,174.87) .. controls (615.5,177.39) and (613.46,179.43) .. (610.94,179.43) -- (547.04,179.43) .. controls (544.53,179.43) and (542.49,177.39) .. (542.49,174.87) -- cycle ;
\draw    (537,77.43) -- (621,77.21) ;
\draw    (537,100.17) -- (580.79,100.17) -- (591.1,100.17) -- (591.2,100.17) -- (620.71,100.17) ;
\draw    (537,123.73) -- (620.71,123.73) ;
\draw    (537,146.84) -- (620.71,146.84) ;
\draw    (537,169.97) -- (620.71,169.97) ;
\draw  [fill={rgb, 255:red, 255; green, 255; blue, 255 }  ,fill opacity=1 ] (561.79,142.82) -- (596.79,142.82) -- (596.79,175.14) -- (561.79,175.14) -- cycle ;
\draw  [color={rgb, 255:red, 0; green, 0; blue, 0 }  ,draw opacity=1 ][dash pattern={on 0.84pt off 2.51pt}] (403,61.57) .. controls (403,57.79) and (406.07,54.72) .. (409.85,54.72) -- (618.35,54.72) .. controls (622.13,54.72) and (625.2,57.79) .. (625.2,61.57) -- (625.2,201.62) .. controls (625.2,205.41) and (622.13,208.47) .. (618.35,208.47) -- (409.85,208.47) .. controls (406.07,208.47) and (403,205.41) .. (403,201.62) -- cycle ;
\draw  [draw opacity=0][fill={rgb, 255:red, 171; green, 207; blue, 207 }  ,fill opacity=1 ] (202.59,86.06) -- (209.65,86.06) -- (209.65,90.22) -- (202.59,90.22) -- cycle ;
\draw  [draw opacity=0][fill={rgb, 255:red, 171; green, 207; blue, 207 }  ,fill opacity=1 ] (314.34,246.66) -- (321.4,246.66) -- (321.4,250.92) -- (314.34,250.92) -- cycle ;
\draw  [draw opacity=0][fill={rgb, 255:red, 171; green, 207; blue, 207 }  ,fill opacity=1 ] (475.8,86.56) -- (482.86,86.56) -- (482.86,90.28) -- (475.8,90.28) -- cycle ;
\draw    (197.23,87.08) .. controls (207.03,81.42) and (206.03,91.88) .. (215.83,85.54) ;
\draw    (197.23,90.8) .. controls (207.03,85.14) and (206.03,95.6) .. (215.83,89.26) ;
\draw    (470.06,87.08) .. controls (479.86,81.42) and (478.86,91.88) .. (488.66,85.54) ;
\draw    (470.06,90.8) .. controls (479.86,85.14) and (478.86,95.6) .. (488.66,89.26) ;
\draw    (308.65,247.52) .. controls (318.45,241.86) and (317.45,252.32) .. (327.25,245.98) ;
\draw    (308.65,251.23) .. controls (318.45,245.57) and (317.45,256.03) .. (327.25,249.69) ;

\draw (8,64.63) node   [align=left] {\begin{minipage}[lt]{11.27pt}\setlength\topsep{0pt}
(a)
\end{minipage}};
\draw (59.19,159.03) node    {$U_{1m}^{( B)}$};
\draw (9.87,224.8) node   [align=left] {\begin{minipage}[lt]{11.27pt}\setlength\topsep{0pt}
(b)
\end{minipage}};
\draw (89.56,319.21) node    {$U_{1m}^{( B)}$};
\draw (49.88,284.11) node    {$P_{B}$};
\draw (129.88,284.5) node    {$P_{B}^{-1}$};
\draw (114.56,64.6) node   [align=left] {\begin{minipage}[lt]{11.44pt}\setlength\topsep{0pt}
(d)
\end{minipage}};
\draw (205.55,158.99) node  [font=\normalsize]  {$U_{m}^{( D_{L})}$};
\draw (152,123.71) node    {$X$};
\draw (152,100.58) node    {$X$};
\draw (152,77.44) node    {$X$};
\draw (263.12,123.71) node    {$X$};
\draw (262.67,100.58) node    {$X$};
\draw (262.67,77.44) node    {$X$};
\draw (62.37,198.08) node  [color={rgb, 255:red, 0; green, 0; blue, 0 }  ,opacity=1 ]  {$I^{\otimes n} \otimes U_{m}^{( B)}$};
\draw (94.18,360.82) node  [color={rgb, 255:red, 0; green, 0; blue, 0 }  ,opacity=1 ]  {$P_{B}\left[ I^{\otimes n} \otimes U_{m}^{( B)}\right] P_{B}^{-1}$};
\draw (207.57,195) node  [color={rgb, 255:red, 0; green, 0; blue, 0 }  ,opacity=1 ]  {$-R_{0}^{( n)} \otimes U_{m}^{( D_{L})}$};
\draw (182,123.76) node    {$H$};
\draw (231,123.76) node    {$H$};
\draw (344.85,158.98) node  [font=\normalsize]  {$U_{m}^{( D_{L})}$};
\draw (346.87,194.99) node  [color={rgb, 255:red, 0; green, 0; blue, 0 }  ,opacity=1 ]  {$I^{\otimes n} \otimes U_{m}^{( D_{L})}$};
\draw (179.83,224.9) node   [align=left] {\begin{minipage}[lt]{13.39pt}\setlength\topsep{0pt}
(c)
\end{minipage}};
\draw (317.35,319.29) node  [font=\normalsize]  {$U_{m}^{( C)}$};
\draw (263.8,284.01) node    {$X$};
\draw (263.8,260.88) node    {$X$};
\draw (263.8,237.74) node    {$X$};
\draw (374.92,284.01) node    {$X$};
\draw (374.47,260.88) node    {$X$};
\draw (374.47,237.74) node    {$X$};
\draw (330.4,356.87) node  [color={rgb, 255:red, 0; green, 0; blue, 0 }  ,opacity=1 ]  {$P_{C}\left[ -R_{0}^{( n)} \otimes U_{m}^{( C)}\right] P_{C}^{-1}$};
\draw (293.8,284.06) node    {$H$};
\draw (342.8,284.06) node    {$H$};
\draw (538.95,319.28) node  [font=\normalsize]  {$U_{m}^{( C)}$};
\draw (540.96,357.86) node  [color={rgb, 255:red, 0; green, 0; blue, 0 }  ,opacity=1 ]  {$P_{C}\left[ I^{\otimes n} \otimes U_{m}^{( C)}\right] P_{C}^{-1}$};
\draw (228.06,283.88) node    {$P_{C}$};
\draw (411.46,284.07) node    {$P_{C}^{-1}$};
\draw (494.43,284.1) node    {$P_{C}$};
\draw (580.43,284.32) node    {$P_{C}^{-1}$};
\draw (413.97,64.6) node   [align=left] {\begin{minipage}[lt]{10.1pt}\setlength\topsep{0pt}
(e)
\end{minipage}};
\draw (478.75,158.99) node  [font=\normalsize]  {$U_{m}^{( D_{R})}$};
\draw (480.77,195) node  [color={rgb, 255:red, 0; green, 0; blue, 0 }  ,opacity=1 ]  {$-R_{1}^{( n)} \otimes U_{m}^{( D_{R})}$};
\draw (455.2,123.76) node    {$H$};
\draw (504.2,123.76) node    {$H$};
\draw (579.29,158.98) node  [font=\normalsize]  {$U_{m}^{( D_{R})}$};
\draw (581.31,194.99) node  [color={rgb, 255:red, 0; green, 0; blue, 0 }  ,opacity=1 ]  {$I^{\otimes n} \otimes U_{m}^{( D_{R})}$};
\draw (344,87.09) node  [font=\normalsize]  {$\vdots $};
\draw (579.06,87.95) node  [font=\normalsize]  {$\vdots $};
\draw (537.73,247.95) node  [font=\normalsize]  {$\vdots $};
\draw (89.5,248.08) node  [font=\normalsize]  {$\vdots $};
\draw (58.5,87.43) node  [font=\normalsize]  {$\vdots $};

\end{tikzpicture}

%% file: block_encoding.tikz
\tikzset{every picture/.style={line width=0.75pt}} 

\begin{tikzpicture}[x=0.75pt,y=0.75pt,yscale=-1,xscale=1]

\draw  [color={rgb, 255:red, 0; green, 0; blue, 0 }  ,draw opacity=1 ][fill={rgb, 255:red, 171; green, 207; blue, 207 }  ,fill opacity=1 ][dash pattern={on 4.5pt off 4.5pt}] (412.04,125.44) .. controls (412.04,123.25) and (413.81,121.47) .. (416,121.47) -- (529.24,121.47) .. controls (531.43,121.47) and (533.21,123.25) .. (533.21,125.44) -- (533.21,226.18) .. controls (533.21,228.37) and (531.43,230.14) .. (529.24,230.14) -- (416,230.14) .. controls (413.81,230.14) and (412.04,228.37) .. (412.04,226.18) -- cycle ;
\draw  [dash pattern={on 0.84pt off 2.51pt}]  (211.5,162.81) -- (332.39,103.2) ;
\draw    (434.09,129.55) -- (434.04,209.44) ;
\draw    (464.25,129.55) -- (464.2,209.45) ;
\draw    (510.59,129.55) -- (510.54,209.45) ;
\draw  [draw opacity=0][fill={rgb, 255:red, 171; green, 207; blue, 207 }  ,fill opacity=1 ] (430.78,160.2) -- (437.35,160.2) -- (437.35,184.49) -- (430.78,184.49) -- cycle ;
\draw  [draw opacity=0][fill={rgb, 255:red, 171; green, 207; blue, 207 }  ,fill opacity=1 ] (460.75,160.2) -- (467.33,160.2) -- (467.33,184.49) -- (460.75,184.49) -- cycle ;
\draw  [draw opacity=0][fill={rgb, 255:red, 171; green, 207; blue, 207 }  ,fill opacity=1 ] (507.35,160.2) -- (513.93,160.2) -- (513.93,184.49) -- (507.35,184.49) -- cycle ;

\draw    (352.62,129.46) -- (582.54,129.46) ;
\draw    (352.62,195.7) -- (582.54,195.7) ;
\draw    (352.6,215.7) -- (582.52,215.7) ;
\draw  [fill={rgb, 255:red, 255; green, 255; blue, 255 }  ,fill opacity=1 ] (431.04,129.55) .. controls (431.04,127.86) and (432.41,126.49) .. (434.09,126.49) .. controls (435.78,126.49) and (437.15,127.86) .. (437.15,129.55) .. controls (437.15,131.23) and (435.78,132.6) .. (434.09,132.6) .. controls (432.41,132.6) and (431.04,131.23) .. (431.04,129.55) -- cycle ;
\draw    (352.6,149.5) -- (582.52,149.5) ;
\draw  [fill={rgb, 255:red, 255; green, 255; blue, 255 }  ,fill opacity=1 ] (431.04,149.55) .. controls (431.04,147.86) and (432.41,146.49) .. (434.09,146.49) .. controls (435.78,146.49) and (437.15,147.86) .. (437.15,149.55) .. controls (437.15,151.23) and (435.78,152.6) .. (434.09,152.6) .. controls (432.41,152.6) and (431.04,151.23) .. (431.04,149.55) -- cycle ;
\draw  [fill={rgb, 255:red, 255; green, 255; blue, 255 }  ,fill opacity=1 ] (431.04,195.55) .. controls (431.04,193.87) and (432.41,192.5) .. (434.09,192.5) .. controls (435.78,192.5) and (437.15,193.87) .. (437.15,195.55) .. controls (437.15,197.24) and (435.78,198.61) .. (434.09,198.61) .. controls (432.41,198.61) and (431.04,197.24) .. (431.04,195.55) -- cycle ;
\draw  [fill={rgb, 255:red, 255; green, 255; blue, 255 }  ,fill opacity=1 ] (373.54,122) .. controls (373.54,122) and (373.54,122) .. (373.54,122) -- (404.04,122) .. controls (404.04,122) and (404.04,122) .. (404.04,122) -- (404.04,202.89) .. controls (404.04,202.89) and (404.04,202.89) .. (404.04,202.89) -- (373.54,202.89) .. controls (373.54,202.89) and (373.54,202.89) .. (373.54,202.89) -- cycle ;
\draw  [fill={rgb, 255:red, 255; green, 255; blue, 255 }  ,fill opacity=1 ] (423,204.7) -- (445,204.7) -- (445,226.7) -- (423,226.7) -- cycle ;
\draw  [fill={rgb, 255:red, 255; green, 255; blue, 255 }  ,fill opacity=1 ] (461.08,129.55) .. controls (461.08,127.87) and (462.45,126.5) .. (464.14,126.5) .. controls (465.83,126.5) and (467.19,127.87) .. (467.19,129.55) .. controls (467.19,131.24) and (465.83,132.61) .. (464.14,132.61) .. controls (462.45,132.61) and (461.08,131.24) .. (461.08,129.55) -- cycle ;
\draw  [fill={rgb, 255:red, 255; green, 255; blue, 255 }  ,fill opacity=1 ] (461.08,149.55) .. controls (461.08,147.87) and (462.45,146.5) .. (464.14,146.5) .. controls (465.83,146.5) and (467.19,147.87) .. (467.19,149.55) .. controls (467.19,151.24) and (465.83,152.61) .. (464.14,152.61) .. controls (462.45,152.61) and (461.08,151.24) .. (461.08,149.55) -- cycle ;
\draw  [fill={rgb, 255:red, 0; green, 0; blue, 0 }  ,fill opacity=1 ] (461.2,195.56) .. controls (461.2,193.88) and (462.56,192.51) .. (464.25,192.51) .. controls (465.94,192.51) and (467.3,193.88) .. (467.3,195.56) .. controls (467.3,197.25) and (465.94,198.62) .. (464.25,198.62) .. controls (462.56,198.62) and (461.2,197.25) .. (461.2,195.56) -- cycle ;
\draw  [fill={rgb, 255:red, 0; green, 0; blue, 0 }  ,fill opacity=1 ] (507.54,129.55) .. controls (507.54,127.87) and (508.91,126.5) .. (510.59,126.5) .. controls (512.28,126.5) and (513.65,127.87) .. (513.65,129.55) .. controls (513.65,131.24) and (512.28,132.61) .. (510.59,132.61) .. controls (508.91,132.61) and (507.54,131.24) .. (507.54,129.55) -- cycle ;
\draw  [fill={rgb, 255:red, 0; green, 0; blue, 0 }  ,fill opacity=1 ] (507.54,149.55) .. controls (507.54,147.87) and (508.91,146.5) .. (510.59,146.5) .. controls (512.28,146.5) and (513.65,147.87) .. (513.65,149.55) .. controls (513.65,151.24) and (512.28,152.61) .. (510.59,152.61) .. controls (508.91,152.61) and (507.54,151.24) .. (507.54,149.55) -- cycle ;
\draw  [fill={rgb, 255:red, 0; green, 0; blue, 0 }  ,fill opacity=1 ] (507.54,195.56) .. controls (507.54,193.88) and (508.91,192.51) .. (510.59,192.51) .. controls (512.28,192.51) and (513.65,193.88) .. (513.65,195.56) .. controls (513.65,197.25) and (512.28,198.62) .. (510.59,198.62) .. controls (508.91,198.62) and (507.54,197.25) .. (507.54,195.56) -- cycle ;
\draw  [fill={rgb, 255:red, 255; green, 255; blue, 255 }  ,fill opacity=1 ] (453.21,204.72) -- (475.21,204.72) -- (475.21,226.72) -- (453.21,226.72) -- cycle ;
\draw  [fill={rgb, 255:red, 255; green, 255; blue, 255 }  ,fill opacity=1 ] (499.52,204.72) -- (521.52,204.72) -- (521.52,226.72) -- (499.52,226.72) -- cycle ;
\draw  [fill={rgb, 255:red, 255; green, 255; blue, 255 }  ,fill opacity=1 ] (543.21,122) .. controls (543.21,122) and (543.21,122) .. (543.21,122) -- (573.81,122) .. controls (573.81,122) and (573.81,122) .. (573.81,122) -- (573.81,202.89) .. controls (573.81,202.89) and (573.81,202.89) .. (573.81,202.89) -- (543.21,202.89) .. controls (543.21,202.89) and (543.21,202.89) .. (543.21,202.89) -- cycle ;
\draw    (89.88,147.4) -- (297.38,147.4) ;
\draw    (89.84,177.4) -- (297.34,177.4) ;
\draw    (89.84,207.43) -- (297.34,207.43) ;
\draw  [fill={rgb, 255:red, 255; green, 255; blue, 255 }  ,fill opacity=1 ] (98.73,136.39) -- (120.73,136.39) -- (120.73,158.39) -- (98.73,158.39) -- cycle ;
\draw  [fill={rgb, 255:red, 0; green, 0; blue, 0 }  ,fill opacity=1 ] (148.76,147.45) .. controls (148.76,145.87) and (150.04,144.59) .. (151.62,144.59) .. controls (153.2,144.59) and (154.48,145.87) .. (154.48,147.45) .. controls (154.48,149.03) and (153.2,150.31) .. (151.62,150.31) .. controls (150.04,150.31) and (148.76,149.03) .. (148.76,147.45) -- cycle ;
\draw  [fill={rgb, 255:red, 0; green, 0; blue, 0 }  ,fill opacity=1 ] (237.3,147.48) .. controls (237.3,145.9) and (238.58,144.62) .. (240.16,144.62) .. controls (241.74,144.62) and (243.02,145.9) .. (243.02,147.48) .. controls (243.02,149.06) and (241.74,150.34) .. (240.16,150.34) .. controls (238.58,150.34) and (237.3,149.06) .. (237.3,147.48) -- cycle ;
\draw    (151.62,147.45) -- (151.63,192.42) ;
\draw    (240.13,147.48) -- (240.14,192.42) ;
\draw  [fill={rgb, 255:red, 255; green, 255; blue, 255 }  ,fill opacity=1 ] (263.98,136.64) -- (285.98,136.64) -- (285.98,158.64) -- (263.98,158.64) -- cycle ;
\draw  [fill={rgb, 255:red, 255; green, 255; blue, 255 }  ,fill opacity=1 ] (136.43,192.22) .. controls (136.43,192.22) and (136.43,192.22) .. (136.43,192.22) -- (166.93,192.22) .. controls (166.93,192.22) and (166.93,192.22) .. (166.93,192.22) -- (166.93,222.5) .. controls (166.93,222.5) and (166.93,222.5) .. (166.93,222.5) -- (136.43,222.5) .. controls (136.43,222.5) and (136.43,222.5) .. (136.43,222.5) -- cycle ;
\draw  [fill={rgb, 255:red, 255; green, 255; blue, 255 }  ,fill opacity=1 ] (225,192.22) .. controls (225,192.22) and (225,192.22) .. (225,192.22) -- (255.5,192.22) .. controls (255.5,192.22) and (255.5,192.22) .. (255.5,192.22) -- (255.5,222.5) .. controls (255.5,222.5) and (255.5,222.5) .. (255.5,222.5) -- (225,222.5) .. controls (225,222.5) and (225,222.5) .. (225,222.5) -- cycle ;
\draw  [dash pattern={on 0.84pt off 2.51pt}]  (211.5,222.81) -- (333.5,250.2) ;
\draw  [fill={rgb, 255:red, 0; green, 0; blue, 0 }  ,fill opacity=1 ] (193,147.46) .. controls (193,145.88) and (194.28,144.6) .. (195.86,144.6) .. controls (197.44,144.6) and (198.72,145.88) .. (198.72,147.46) .. controls (198.72,149.04) and (197.44,150.32) .. (195.86,150.32) .. controls (194.28,150.32) and (193,149.04) .. (193,147.46) -- cycle ;
\draw    (195.86,147.46) -- (195.87,162.4) ;
\draw  [fill={rgb, 255:red, 255; green, 255; blue, 255 }  ,fill opacity=1 ] (181,162.81) .. controls (181,162.81) and (181,162.81) .. (181,162.81) -- (211.5,162.81) .. controls (211.5,162.81) and (211.5,162.81) .. (211.5,162.81) -- (211.5,222.81) .. controls (211.5,222.81) and (211.5,222.81) .. (211.5,222.81) -- (181,222.81) .. controls (181,222.81) and (181,222.81) .. (181,222.81) -- cycle ;
\draw  [dash pattern={on 0.84pt off 2.51pt}] (327.75,111.14) .. controls (327.75,106.03) and (331.89,101.89) .. (337,101.89) -- (584.08,101.89) .. controls (589.19,101.89) and (593.33,106.03) .. (593.33,111.14) -- (593.33,241.64) .. controls (593.33,246.75) and (589.19,250.89) .. (584.08,250.89) -- (337,250.89) .. controls (331.89,250.89) and (327.75,246.75) .. (327.75,241.64) -- cycle ;

\draw (361.55,168.83) node    {$\vdots $};
\draw (388.79,162.45) node    {$U_{\alpha }$};
\draw (434,215.7) node    {$U_{1}$};
\draw (464.21,215.72) node    {$U_{2}$};
\draw (510.52,215.72) node    {$U_{M}$};
\draw (558.51,162.45) node    {$U_{\alpha }^{\dagger }$};
\draw (487.8,172.03) node    {$\cdots $};
\draw (474.89,241.52) node    {$U_{s}$};
\draw (434.07,170.39) node    {$\vdots $};
\draw (464.04,170.39) node    {$\vdots $};
\draw (510.64,170.39) node    {$\vdots $};
\draw (109.73,147.39) node    {$H$};
\draw (151.68,207.36) node    {$U_{\phi }$};
\draw (274.98,147.64) node    {$H$};
\draw (240.25,207.36) node    {$U_{f}^{\dagger }$};
\draw (196.25,192.81) node    {$U$};
\draw (340,113.5) node   [align=left] {(b)};
\draw (57,128.5) node   [align=left] {(a)};
\draw (87.88,147.4) node [anchor=east] [inner sep=0.75pt]    {$\ket{0}_{\mathrm{a}}$};
\draw (87.84,177.4) node [anchor=east] [inner sep=0.75pt]    {$\ket{0}_{\mathrm{s}}$};
\draw (87.84,207.43) node [anchor=east] [inner sep=0.75pt]    {$\ket{0}_{\mathrm{w}}$};

\end{tikzpicture}

%% file: result.tikz
\tikzset{every picture/.style={line width=0.75pt}} 

\begin{tikzpicture}[x=0.75pt,y=0.75pt,yscale=-1,xscale=1]

\draw (548.07,174.75) node  {\includegraphics[width=114.1pt,height=72.96pt]{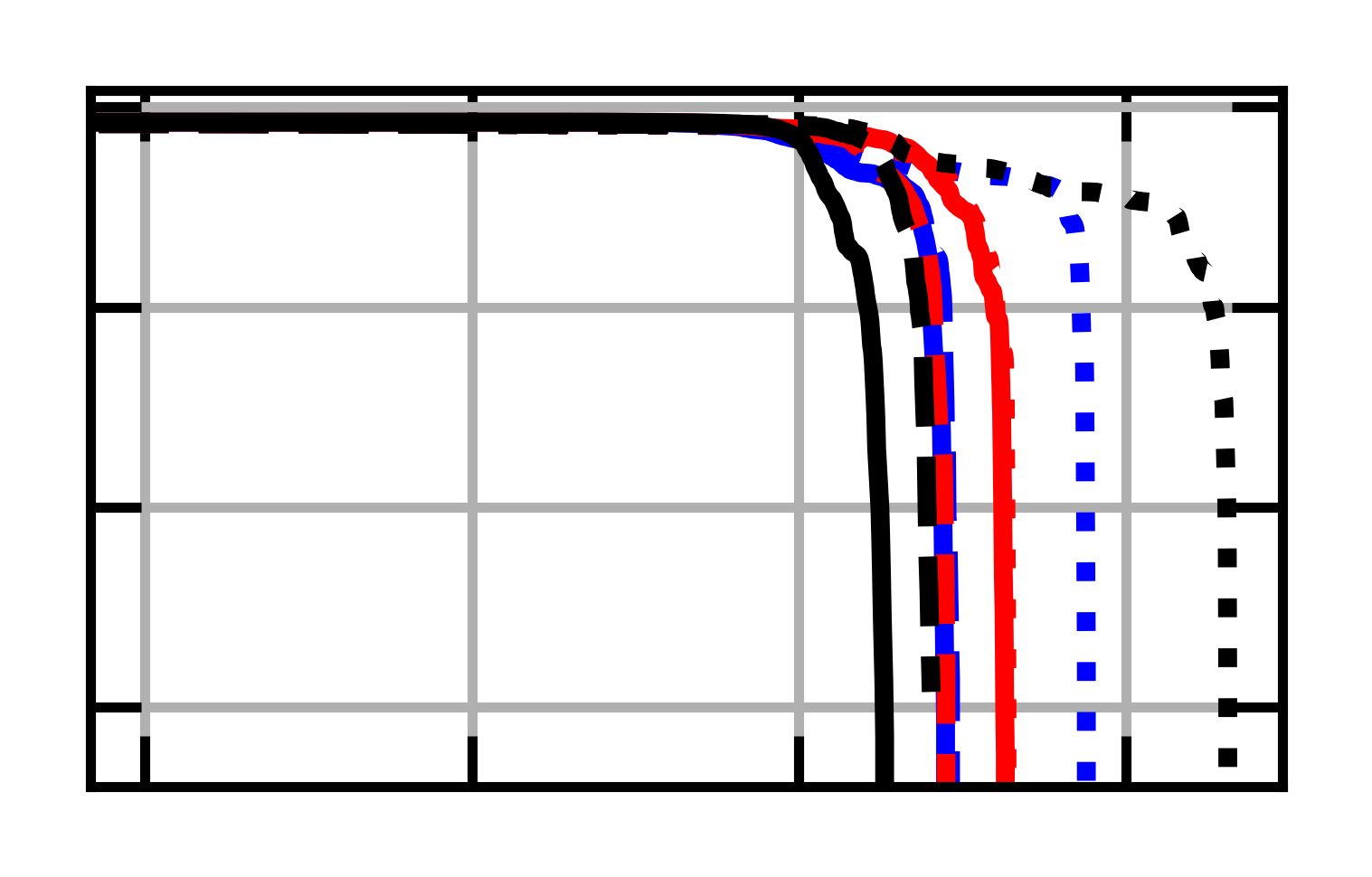}};
\draw (408.07,174.75) node  {\includegraphics[width=114.1pt,height=72.96pt]{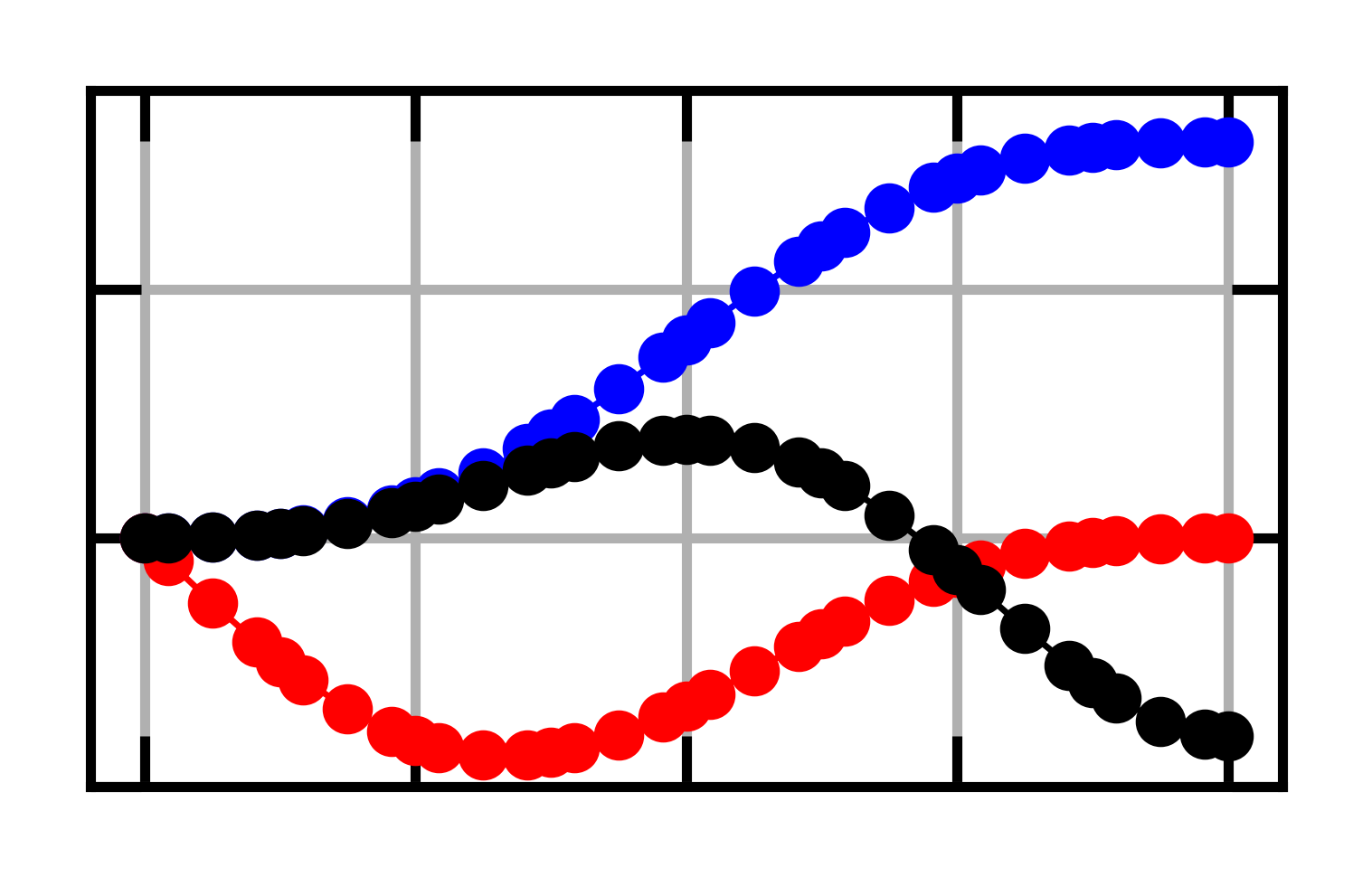}};
\draw (268.07,174.75) node  {\includegraphics[width=114.1pt,height=72.96pt]{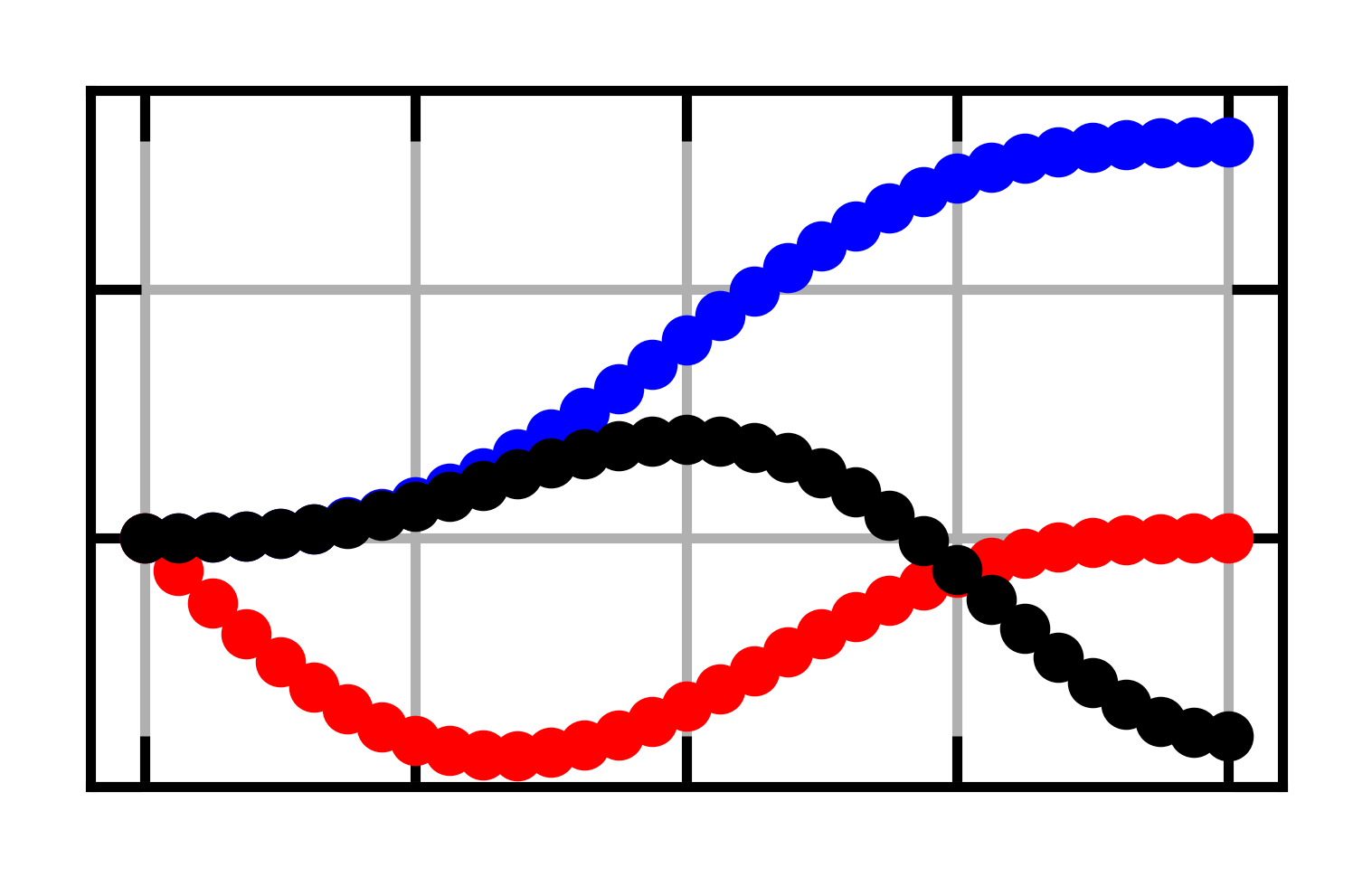}};
\draw (128.07,174.75) node  {\includegraphics[width=114.1pt,height=72.96pt]{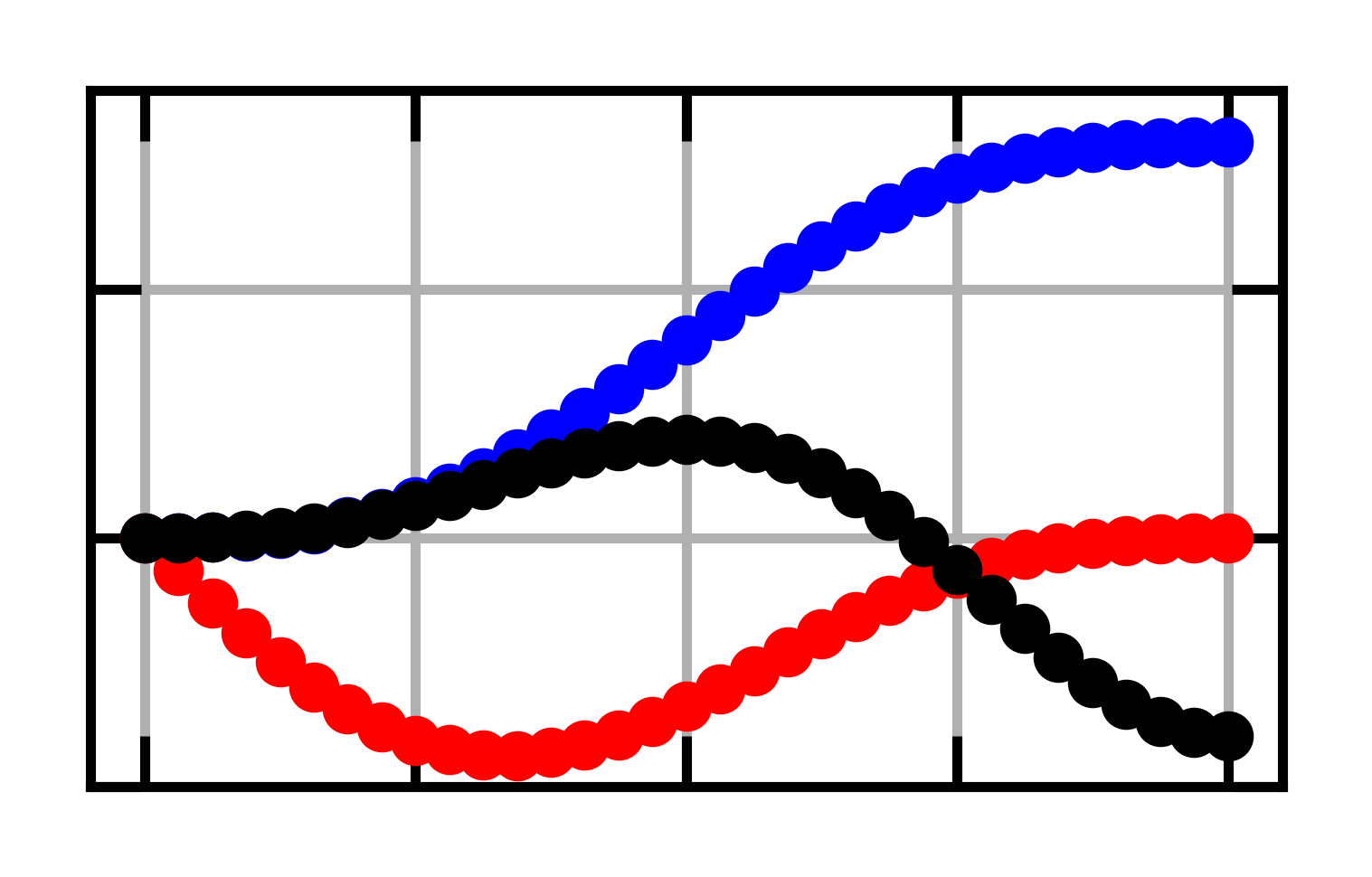}};
\draw  [color={rgb, 255:red, 0; green, 0; blue, 255 }  ,draw opacity=1 ][fill={rgb, 255:red, 0; green, 0; blue, 255 }  ,fill opacity=1 ] (256.33,121.78) .. controls (256.33,120.13) and (257.68,118.78) .. (259.33,118.78) .. controls (260.99,118.78) and (262.33,120.13) .. (262.33,121.78) .. controls (262.33,123.44) and (260.99,124.78) .. (259.33,124.78) .. controls (257.68,124.78) and (256.33,123.44) .. (256.33,121.78) -- cycle ;
\draw  [color={rgb, 255:red, 255; green, 0; blue, 0 }  ,draw opacity=1 ][fill={rgb, 255:red, 255; green, 0; blue, 0 }  ,fill opacity=1 ] (313.67,121.8) .. controls (313.67,120.14) and (315.01,118.8) .. (316.67,118.8) .. controls (318.32,118.8) and (319.67,120.14) .. (319.67,121.8) .. controls (319.67,123.46) and (318.32,124.8) .. (316.67,124.8) .. controls (315.01,124.8) and (313.67,123.46) .. (313.67,121.8) -- cycle ;
\draw  [color={rgb, 255:red, 0; green, 0; blue, 0 }  ,draw opacity=1 ][fill={rgb, 255:red, 0; green, 0; blue, 0 }  ,fill opacity=1 ] (371,121.8) .. controls (371,120.14) and (372.34,118.8) .. (374,118.8) .. controls (375.66,118.8) and (377,120.14) .. (377,121.8) .. controls (377,123.46) and (375.66,124.8) .. (374,124.8) .. controls (372.34,124.8) and (371,123.46) .. (371,121.8) -- cycle ;
\draw [line width=1.5]    (491.57,170.33) -- (511.57,170.33) ;
\draw [line width=1.5]  [dash pattern={on 5.63pt off 4.5pt}]  (491.57,182.4) -- (511.57,182.4) ;
\draw [line width=1.5]  [dash pattern={on 1.69pt off 2.76pt}]  (491.57,194.47) -- (511.57,194.47) ;
\draw   (252.33,113.47) -- (426.67,113.47) -- (426.67,130.14) -- (252.33,130.14) -- cycle ;

\draw (67.46,139.53) node [anchor=north west][inner sep=0.75pt]   [align=left] {(a)};
\draw (208.83,139.23) node [anchor=north west][inner sep=0.75pt]   [align=left] {(b)};
\draw (349.4,139.37) node [anchor=north west][inner sep=0.75pt]   [align=left] {(c)};
\draw (68.52,214.27) node [anchor=north] [inner sep=0.75pt]    {$0$};
\draw (97.5,214.17) node [anchor=north] [inner sep=0.75pt]    {$0.25$};
\draw (128.88,214.17) node [anchor=north] [inner sep=0.75pt]    {$0.5$};
\draw (157.86,214.07) node [anchor=north] [inner sep=0.75pt]    {$0.75$};
\draw (189.05,214.07) node [anchor=north] [inner sep=0.75pt]    {$1$};
\draw (128.44,226.4) node [anchor=north] [inner sep=0.75pt]    {$x$};
\draw (61.11,184.73) node [anchor=east] [inner sep=0.75pt]    {$0$};
\draw (62.1,158.16) node [anchor=east] [inner sep=0.75pt]    {$0.1$};
\draw (36.5,176.86) node [anchor=south] [inner sep=0.75pt]  [rotate=-270]  {$\phi ( x)$};
\draw (264.33,121.78) node [anchor=west] [inner sep=0.75pt]    {$k=0$};
\draw (321.8,121.5) node [anchor=west] [inner sep=0.75pt]    {$k=\pi $};
\draw (379,121.5) node [anchor=west] [inner sep=0.75pt]    {$k=2\pi $};
\draw (210,214.27) node [anchor=north] [inner sep=0.75pt]    {$0$};
\draw (238.98,214.17) node [anchor=north] [inner sep=0.75pt]    {$0.25$};
\draw (270.36,214.17) node [anchor=north] [inner sep=0.75pt]    {$0.5$};
\draw (299.34,214.07) node [anchor=north] [inner sep=0.75pt]    {$0.75$};
\draw (330.53,214.07) node [anchor=north] [inner sep=0.75pt]    {$1$};
\draw (269.92,226.4) node [anchor=north] [inner sep=0.75pt]    {$x$};
\draw (349,214.27) node [anchor=north] [inner sep=0.75pt]    {$0$};
\draw (377.98,214.17) node [anchor=north] [inner sep=0.75pt]    {$0.25$};
\draw (409.36,214.17) node [anchor=north] [inner sep=0.75pt]    {$0.5$};
\draw (438.34,214.07) node [anchor=north] [inner sep=0.75pt]    {$0.75$};
\draw (469.53,214.07) node [anchor=north] [inner sep=0.75pt]    {$1$};
\draw (408.92,226.4) node [anchor=north] [inner sep=0.75pt]    {$x$};
\draw (486.73,141.37) node [anchor=north west][inner sep=0.75pt]   [align=left] {(d)};
\draw (514.54,165.37) node [anchor=west] [inner sep=0.75pt]    {$p=1$};
\draw (513.87,179.18) node [anchor=west] [inner sep=0.75pt]    {$p=2$};
\draw (513.73,193) node [anchor=west] [inner sep=0.75pt]    {$p=4$};
\draw (490.84,214.27) node [anchor=north] [inner sep=0.75pt]    {$1$};
\draw (524.75,214.17) node [anchor=north] [inner sep=0.75pt]    {$10$};
\draw (561.48,214.17) node [anchor=north] [inner sep=0.75pt]    {$100$};
\draw (597.79,214.17) node [anchor=north] [inner sep=0.75pt]    {$1000$};
\draw (549.23,225.37) node [anchor=north] [inner sep=0.75pt]    {$\mathrm{Iteration}$};
\draw (650.1,176.54) node [anchor=south] [inner sep=0.75pt]  [rotate=-90]  {$||A\boldsymbol{\phi } -\boldsymbol{f} ||^{2}$};
\draw (615.23,160.52) node [anchor=west] [inner sep=0.75pt]    {$10^{-4}$};
\draw (615.23,181.24) node [anchor=west] [inner sep=0.75pt]    {$10^{-8}$};
\draw (615.23,201.54) node [anchor=west] [inner sep=0.75pt]    {$10^{-12}$};
\draw (615.2,139.07) node [anchor=west] [inner sep=0.75pt]    {$10^{0}$};

\end{tikzpicture}

%% file: kldiv.tikz
  
\tikzset {_qpbj4m0wn/.code = {\pgfsetadditionalshadetransform{ \pgftransformshift{\pgfpoint{0 bp } { 0 bp }  }  \pgftransformrotate{-270 }  \pgftransformscale{2 }  }}}
\pgfdeclarehorizontalshading{_v2q1aits0}{150bp}{rgb(0bp)=(0.78,0.86,0.94);
rgb(37.5bp)=(0.78,0.86,0.94);
rgb(49.64285714285714bp)=(0.26,0.57,0.78);
rgb(62.5bp)=(0.03,0.19,0.42);
rgb(100bp)=(0.03,0.19,0.42)}
\tikzset{every picture/.style={line width=0.75pt}} 

\begin{tikzpicture}[x=0.75pt,y=0.75pt,yscale=-1,xscale=1]

\draw (292.67,129.75) node  {\includegraphics[width=102pt,height=116.99pt]{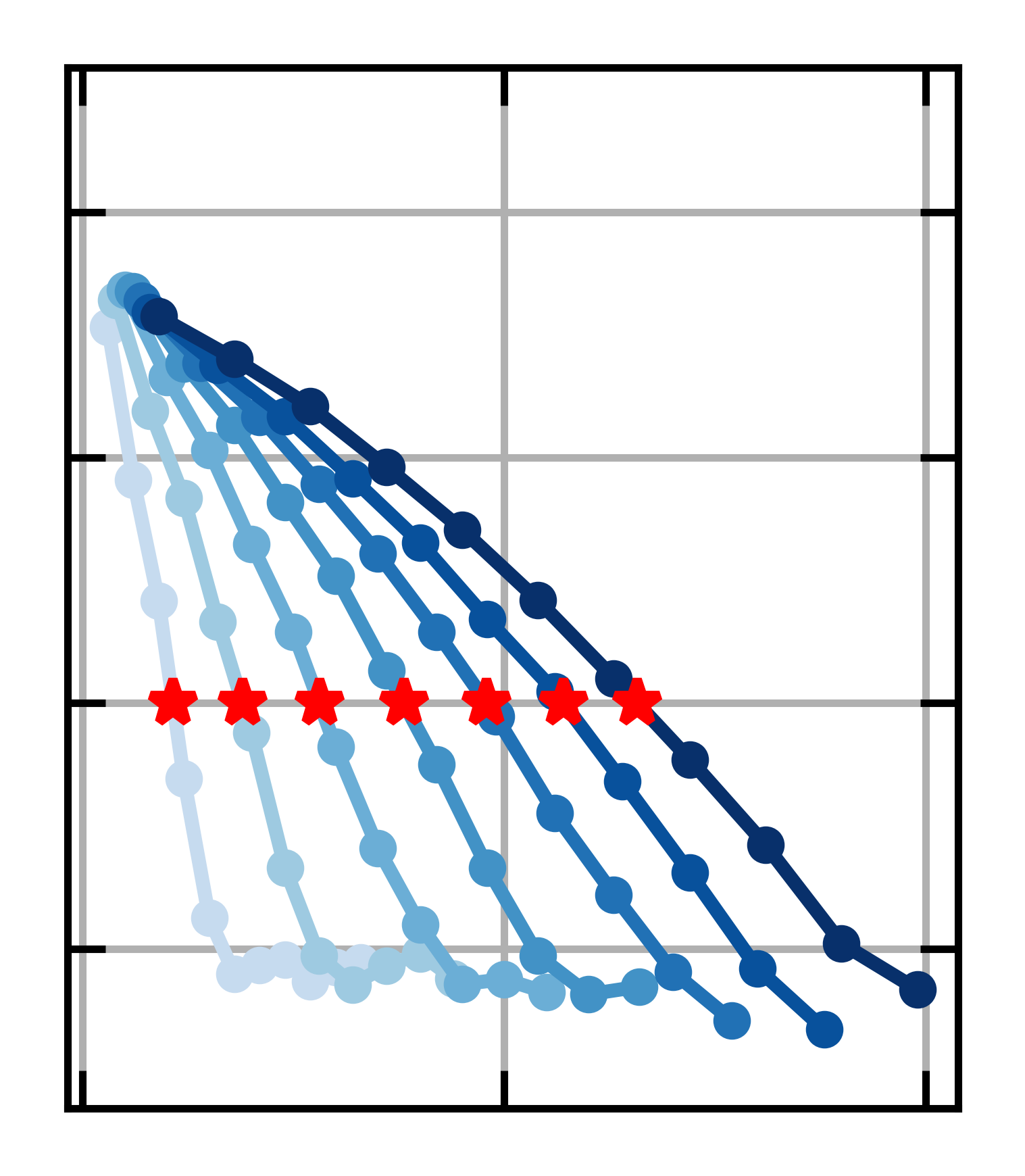}};
\draw (156.67,129.75) node  {\includegraphics[width=102pt,height=116.99pt]{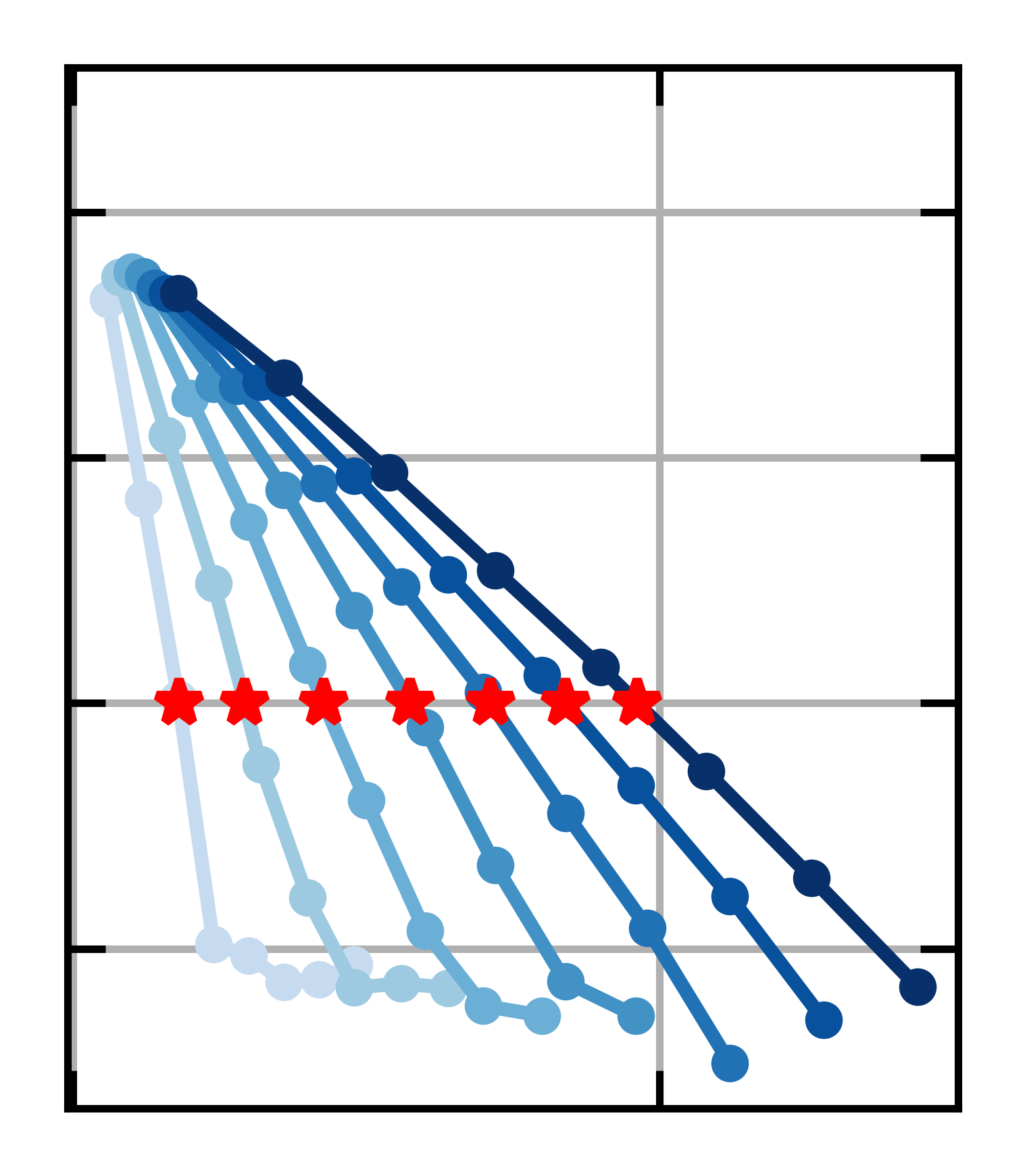}};
\draw  [draw opacity=0][fill={rgb, 255:red, 241; green, 241; blue, 241 }  ,fill opacity=1 ] (158.13,62.47) -- (213.73,62.47) -- (213.73,118.01) -- (158.13,118.01) -- cycle ;
\draw  [draw opacity=0][fill={rgb, 255:red, 241; green, 241; blue, 241 }  ,fill opacity=1 ] (294.13,62.22) -- (349.73,62.22) -- (349.73,117.76) -- (294.13,117.76) -- cycle ;
\draw [color={rgb, 255:red, 0; green, 0; blue, 0 }  ,draw opacity=1 ] [dash pattern={on 4.5pt off 4.5pt}]  (158.53,107.09) -- (211.37,68.46) ;
\draw [color={rgb, 255:red, 0; green, 0; blue, 0 }  ,draw opacity=1 ] [dash pattern={on 4.5pt off 4.5pt}]  (294.12,105.85) -- (347.1,64.64) ;
\draw [color={rgb, 255:red, 255; green, 0; blue, 0 }  ,draw opacity=1 ] [dash pattern={on 4.5pt off 4.5pt}]  (98,145.1) -- (215.78,145.14) ;
\draw [color={rgb, 255:red, 255; green, 0; blue, 0 }  ,draw opacity=1 ] [dash pattern={on 4.5pt off 4.5pt}]  (233.78,145.1) -- (351.56,145.14) ;
\path  [shading=_v2q1aits0,_qpbj4m0wn] (360.94,72) -- (366.61,72) -- (366.61,192) -- (360.94,192) -- cycle ; 
 \draw   (360.94,72) -- (366.61,72) -- (366.61,192) -- (360.94,192) -- cycle ; 

\draw    (360.94,72) -- (366.61,72) ;
\draw    (360.94,192) -- (366.61,192) ;
\draw    (360.9,92) -- (366.57,92) ;
\draw    (360.9,112) -- (366.57,112) ;
\draw    (360.9,132) -- (366.57,132) ;
\draw    (360.9,152) -- (366.57,152) ;
\draw    (360.9,172) -- (366.57,172) ;
\draw (186.1,90.27) node  {\includegraphics[width=52.5pt,height=52.5pt]{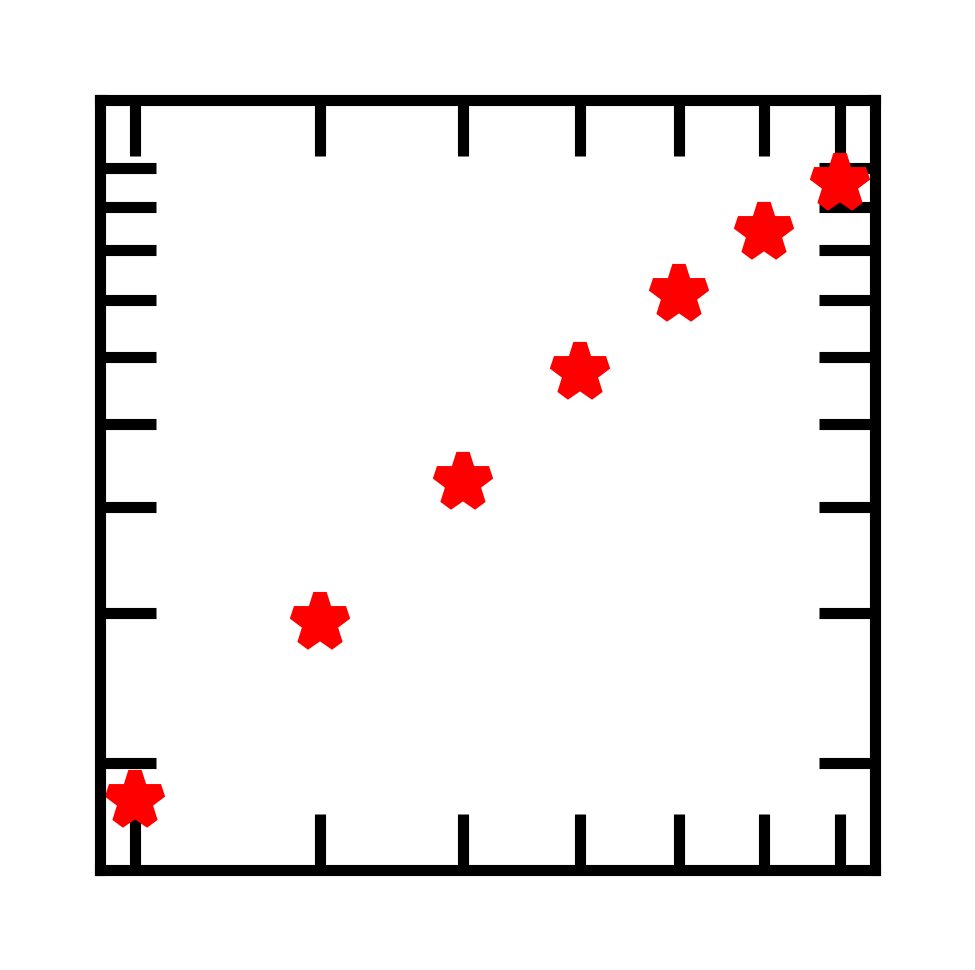}};
\draw (321.9,89.74) node  {\includegraphics[width=52.5pt,height=52.5pt]{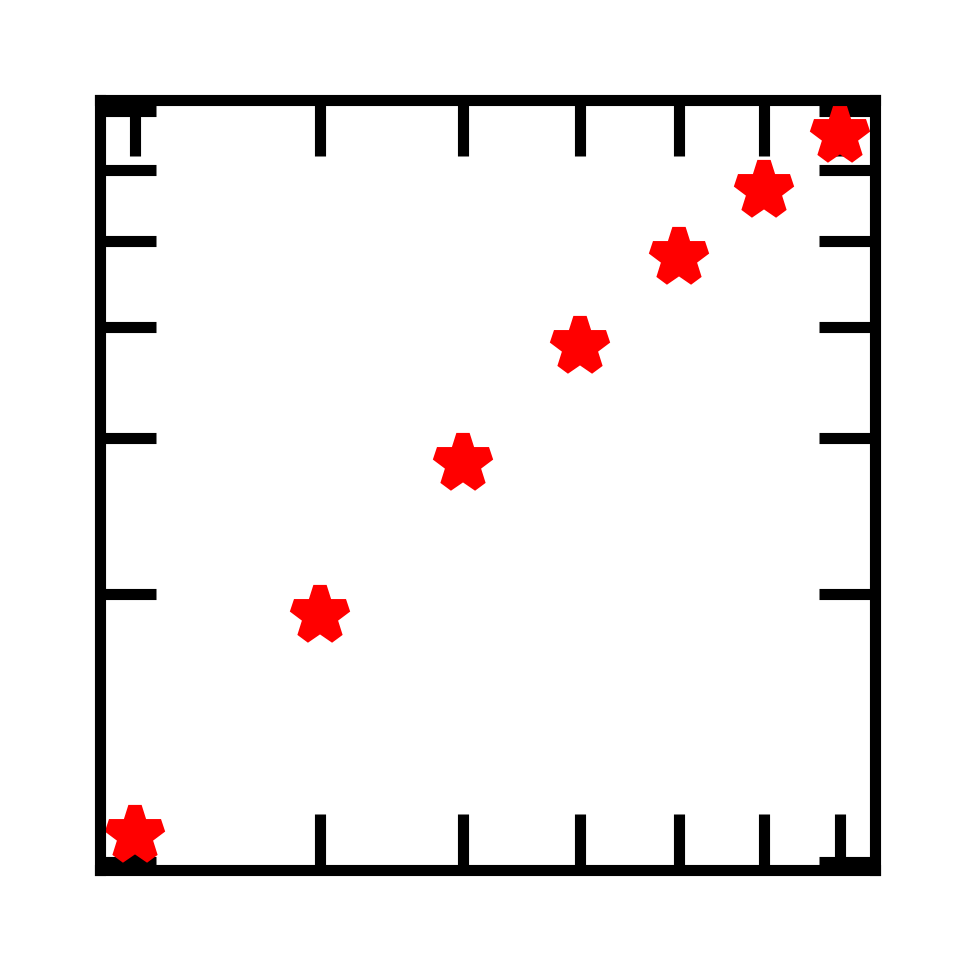}};

\draw (99.5,62) node [anchor=north west][inner sep=0.75pt]   [align=left] {(a)};
\draw (235.5,62) node [anchor=north west][inner sep=0.75pt]   [align=left] {(b)};
\draw (237.22,198.65) node [anchor=north] [inner sep=0.75pt]    {$0$};
\draw (292,198.7) node [anchor=north] [inner sep=0.75pt]    {$50$};
\draw (347,198.7) node [anchor=north] [inner sep=0.75pt]    {$100$};
\draw (100.48,198.7) node [anchor=north] [inner sep=0.75pt]    {$0$};
\draw (176,198.75) node [anchor=north] [inner sep=0.75pt]    {$100$};
\draw (98.1,178.14) node [anchor=east] [inner sep=0.75pt]    {$10^{-6}$};
\draw (98.1,144) node [anchor=east] [inner sep=0.75pt]    {$10^{-4}$};
\draw (98.1,111.2) node [anchor=east] [inner sep=0.75pt]    {$10^{-2}$};
\draw (98.1,78) node [anchor=east] [inner sep=0.75pt]    {$10^{0}$};
\draw (150.58,216.4) node    {$\dim(\boldsymbol{\theta })$};
\draw (295,215.6) node    {$\dim(\boldsymbol{\theta })$};
\draw (51.83,192.37) node [anchor=north west][inner sep=0.75pt]  [rotate=-270]  {$D_{\mathrm{KL}}( \rho ( U_{\phi }) ||\rho _{\mathrm{Haar}})$};
\draw (186.37,118.61) node [anchor=north] [inner sep=0.75pt]  [font=\scriptsize]  {$n$};
\draw (151.96,84.51) node [anchor=south] [inner sep=0.75pt]  [font=\scriptsize,rotate=-270]  {$\dim(\boldsymbol{\theta })$};
\draw (324.65,118.04) node [anchor=north] [inner sep=0.75pt]  [font=\scriptsize]  {$n$};
\draw (288.3,83.93) node [anchor=south] [inner sep=0.75pt]  [font=\scriptsize,rotate=-270]  {$\dim(\boldsymbol{\theta })$};
\draw (368.61,192) node [anchor=west] [inner sep=0.75pt]  [font=\scriptsize]  {$3$};
\draw (368.57,172) node [anchor=west] [inner sep=0.75pt]  [font=\scriptsize]  {$4$};
\draw (368.57,152) node [anchor=west] [inner sep=0.75pt]  [font=\scriptsize]  {$5$};
\draw (368.57,132) node [anchor=west] [inner sep=0.75pt]  [font=\scriptsize]  {$6$};
\draw (368.57,112) node [anchor=west] [inner sep=0.75pt]  [font=\scriptsize]  {$7$};
\draw (368.57,92) node [anchor=west] [inner sep=0.75pt]  [font=\scriptsize]  {$8$};
\draw (368.61,72) node [anchor=west] [inner sep=0.75pt]  [font=\scriptsize]  {$9$};
\draw (380.25,126.01) node  [font=\normalsize]  {$n$};
\draw (160.19,56.69) node [anchor=south] [inner sep=0.75pt]    {$SU\left( 2^{n}\right)$};
\draw (295.53,56.66) node [anchor=south] [inner sep=0.75pt]    {$SO\left( 2^{n}\right)$};
\draw (180.72,80.79) node  [font=\scriptsize,rotate=-326.13]  {$\varpropto n^{1.25}$};
\draw (316.72,78.91) node  [font=\scriptsize,rotate=-326.13]  {$\varpropto n^{1.25}$};

\end{tikzpicture}